\documentclass[12pt]{article}
\usepackage{epsfig}
\usepackage{amsmath}
\usepackage{youngtab}

\newcommand{\tr}{\mathop{\rm tr}}

\newcommand{\be}{\begin{equation}}
\newcommand{\ee}{\end{equation}}
\newcommand{\bea}{\begin{eqnarray}}
\newcommand{\eea}{\end{eqnarray}}
\newcommand{\ap}{\alpha^\prime}

\newcommand{\nn}{\nonumber}
\newcommand{\Rhat}{\widehat R}

\newcommand{\labphi}[2]{\phi^{\,\Yboxdim5pt\tiny\young(#1,#2)}}

\newcommand{\suphi}{\phi^{\,\Yboxdim5pt\yng(1,1)}}
\newcommand{\suchi}{\chi^{\,\Yboxdim5pt\yng(1)}}
\newcommand{\suchib}{\overline\chi^{\,\Yboxdim5pt\yng(1,1,1)}}

\def\bra#1{\left<#1\right|}
\def\ket#1{\left|#1\right>}
{\catcode`\|=\active
  \gdef\Braket#1{\left<\mathcode`\|"8000\let|\bravert {#1}\right>}}
\def\bravert{\egroup\,\vrule\,\bgroup}
\begin{document}
\begin{titlepage}

\begin{flushright}
PUPT-2089\\ CALT-68-2426\\ hep-th/0307032
\end{flushright}
\vspace{8 mm}

\begin{center}
{\huge Quantizing String Theory in $AdS_5\times S^5$:
Beyond the pp-Wave}
\end{center}

\vspace{8 mm}

\begin{center}
{\large Curtis G.\ Callan, Jr.${}^{a}$,
Hok Kong Lee${}^{b}$, Tristan McLoughlin${}^{b}$, \\
John H. Schwarz${}^{b}$,
Ian Swanson${}^{b}$, and Xinkai Wu${}^{b}$ }\\

\vspace{3mm}

${}^a$ Joseph Henry Laboratories\\
Princeton University\\
Princeton, New Jersey 08544, USA\\
\vspace{0.5 cm}
${}^b$ California Institute of Technology\\
Pasadena, CA 91125, USA
\end{center}

\vspace{5 mm}

\begin{center}
{\large Abstract}
\end{center}

In a certain kinematic limit, where the effects of spacetime curvature
(and other background fields) greatly simplify, the light-cone gauge
world-sheet action for a type IIB superstring
on $AdS_5\times S^5$ reduces to that
of a free field theory. It has been conjectured by Berenstein, Maldacena, and
Nastase that the energy spectrum of this string theory matches the
dimensions of operators in the appropriately defined
large $R$-charge large-$N_c$ sector of ${\cal N}=4$
supersymmetric Yang--Mills theory in four dimensions.
This holographic equivalence is thought to be exact, independent of
any simplifying kinematic limits. As a step toward verifying this larger
conjecture, we have computed the complete set of first curvature
corrections to the spectrum of light-cone gauge
string theory that arises in the expansion of $AdS_5\times S^5$
about the plane-wave limit. 
The resulting spectrum has the complete dependence on 
$\lambda = g_{YM}^2 N_c$; corresponding results in the 
gauge theory are known only to second order in $\lambda$.
We find precise agreement to this order, including the ${\cal N}=4$ extended 
supermultiplet structure.
In the process, we demonstrate that the complicated schemes put forward
in recent years for defining the Green--Schwarz superstring action in
background Ramond-Ramond fields can be reduced to a practical (and correct)
method for quantizing the string.

\vspace{.5cm}
\begin{flushleft}
\today
\end{flushleft}

\end{titlepage}
\newpage

\section{Introduction}
A dramatic prediction of the AdS/CFT correspondence \cite{Maldacena:1997re}
is that the excited state energies of the first-quantized type IIB
superstring in the $AdS_5\times S^5$ geometry should match the dimensions of
certain operators in the strong coupling limit of ${\cal N}=4$
super Yang--Mills field theory in four dimensions \cite{Witten:1998qj}.
The obstacles to
verifying this conjecture are, on one hand, the difficulty of
quantizing superstring theory in the presence of Ramond-Ramond (RR) fields (an
essential feature of the $AdS_5\times S^5$ background) and, on the other,
the need to calculate dimensions of non-BPS operators in strongly coupled
gauge theory.  Substantial progress in both regards has been made recently.

The first step was the realization that the Green--Schwarz (GS)
superstring, evaluated in the light-cone gauge, becomes a free
(albeit massive) worldsheet theory if the true $AdS_5\times S^5$
background is replaced by a Penrose limit describing the near
neighborhood of an equatorial lightlike geodesic on the $S^5$
subspace \cite{Blau:2001ne, Metsaev:2001bj}. The energy spectrum
of this free theory is simply that of a string moving around the
equator of the $S^5$ and boosted to large angular momentum $J$. By
the AdS/CFT correspondence, these string energies should match the
dimensions of operators with large $R$-charge ($R\sim J$) in
strongly coupled four-dimensional ${\cal N} = 4$ super Yang--Mills
theory. This general expectation was precisely realized by
Berenstein, Maldacena, and Nastase (BMN) \cite{Berenstein:2002jq}
who identified the subspace of gauge theory operators
corresponding to specific free string excited states (i.e., states
with different numbers and types of string oscillator modes
applied to the string ground state) and showed that perturbative
calculations of the dimensions of these operators are reliable in
the large $R$-charge limit and gave evidence that they agree with
the string theory predictions. This work demonstrated an
equivalence between energies of string excited states and gauge
theory operator dimensions in a kinematic limit where the
difficulties of quantizing the string and computing operator
dimensions are neatly circumvented.

This equivalence should not be restricted to the Penrose limit of
the geometry (or the large $R$-charge limit of operator
dimensions). It is not easy to verify this stronger prediction,
primarily because the superstring propagating in the general
$AdS_5\times S^5$ geometry is governed by a complicated
interacting worldsheet theory. However, these interactions vanish
in the limit of large $S^5$ angular momentum $J$ and it should be
possible to develop a perturbative expansion (in inverse powers of
$J$) of the string energy spectrum. Corrections to the string
spectrum should subsequently be compared with an expansion in
inverse powers of $R$-charge of the dimensions of BMN-type gauge
theory operators. For various reasons, a lot of attention has been
paid to the problem of calculating these operator dimensions, and
there is an extensive literature ranging over many topics: finding
the proper limit to take to see the correspondence \cite{GenusCounting},
careful calculations of operator dimensions at one loop \cite{OpDimCalcs}, 
making explicit the extended supersymmetry structure of the problem
\cite{BeisertSUSY}, and, more recently, calculations of higher-loop
anomalous dimensions \cite{Beisert:2003tq} (and this is by no means an
all-inclusive list of relevant papers!). The problem of developing the
perturbation theory of worldsheet string dynamics in the
$AdS_5\times S^5$ background has, however, received much less
attention. There has been one study along these lines of the {\it
bosonic} string \cite{Parnachev:2002kk} with promising results
which do not, however, address the crucial supersymmetry issues.
The perturbative analysis of the full GS superstring in the
$AdS_5\times S^5$ background remains to be done, and that is the
subject of this paper. We present here a condensed summary of our
findings, relegating the many cumbersome technical details to a
longer paper \cite{CGCIanTristan}.

In brief, we find that string energies organize themselves into
supermultiplets that match the dimensions of BMN operators in ${\cal N}=4$
super Yang--Mills theory, thus verifying the AdS/CFT correspondence in a
new and challenging context. To achieve this, we have had to perform a
completely explicit quantization of the interacting superstring in a
RR background.  A side benefit of our results is therefore a
verification of the practical utility (and correctness) of the rather
complicated nonlinear action that has been proposed for the fermionic
degrees of freedom of the type IIB GS superstring \cite{Green:1983wt}
in the $AdS_5\times S^5$ background \cite{MetTseyt,Kallosh:1998zx}.
\section{Setup, Notation, Recap of BMN}

We begin with a brief review of essential results from recent
literature on the AdS/CFT correspondence in the Penrose limit.
In convenient global coordinates, the ${ AdS}_5 \times S^5$
metric can be written in the form
\begin{equation}
\label{adsmetric}
ds^2 = \widehat R^2 ( - {\rm cosh}^2 \rho~ dt^2 + d \rho^2 + {\rm sinh}^2
\rho~ d \Omega_3^2 + {\rm cos}^2  \theta~ d \phi^2 +  d \theta^2 +
{\rm sin}^2 \theta~ d \widetilde\Omega_3^2)~,
\end{equation}
where $\widehat R$ denotes the radius of both the sphere and the
AdS space. (The hat is introduced because we reserve the symbol
$R$ for $R$-charge.) The coordinate $\phi$ is periodic with period
$2\pi$ and, strictly speaking, the time coordinate $t$ exhibits
the same periodicity. In order to accommodate string dynamics, it
is necessary to pass to the covering space in which time is {\sl
not} taken to be periodic. This geometry is accompanied by an RR
field with $N_c$ units of flux on the $S^5$. It is a consistent,
maximally supersymmetric type IIB superstring background provided
that 
\begin{equation}
\widehat R^4 =  g_s N_c (\alpha^{\prime})^2~,
\end{equation}
where $g_s$ is the string coupling. The AdS/CFT correspondence asserts that this
string theory is equivalent to ${\cal N} =4$ super Yang--Mills
theory in four dimensions with an $SU(N_c)$ gauge group and coupling
constant $ g_{YM}^2 = g_s $. To simplify both sides of the
correspondence, we study the duality in the simultaneous limits
$g_s\to 0$ (the classical limit of the string theory) and
$N_c\to\infty$ (the planar diagram limit of the gauge theory) with
the 't Hooft coupling $g^2_{YM}N_c$ held fixed. The holographically
dual gauge theory is defined on the conformal boundary of
$AdS_5\times S^5$, which, in this case, is $R\times S^3$.  Duality
demands that operator dimensions in the conformally invariant
gauge theory be equal to the energies of corresponding states of
the `first-quantized' string propagating in the $AdS_5\times S^5$
background. This conjecture is motivated by the fact that both
theories are invariant under the same supergroup $PSU(2,2|4)$, but
a vast amount of more specific evidence in support of the AdS/CFT
correspondence has been accumulated.

As explained above, the quantization problem is simplified by
boosting the string to lightlike momentum along some direction or,
equivalently, by quantizing the string in the background obtained
by taking a Penrose limit of the original geometry using the
lightlike geodesic corresponding to the boosted trajectory. The
simplest choice is to boost along an equator of the $S^5$ or,
equivalently, to take a Penrose limit with respect to the
lightlike geodesic $\phi=t,~\rho=\theta=0$. To perform light-cone
quantization about this geodesic, it is helpful to make the
reparametrizations
\begin{eqnarray}
    \cosh\rho  =  \frac{1+z^2/4}{1-z^2/4} \qquad
    \cos\theta  =  \frac{1-y^2/4}{1+ y^2/4}\ ,
\end{eqnarray}
and work with the metric
\begin{equation}
\label{metric}
ds^2  = \widehat R^2
\biggl[ -\left({1+ \frac{1}{4}z^2\over 1-\frac{1}{4}z^2}\right)^2dt^2
        +\left({1-\frac{1}{4}y^2\over 1+\frac{1}{4}y^2}\right)^2d\phi^2
    + \frac{dz_k dz_k}{(1-\frac{1}{4}z^2)^{2}}
    + \frac{dy_{k'} dy_{k'}}{(1+\frac{1}{4}y^2)^{2}} \biggr]\, ,
\end{equation}
where $y^2 = y_{k'} y^{k'}$ with $k'=5,\dots,8$ and $z^2 = z_k
z^k$ with $k=1,\dots,4$ define eight `Cartesian' coordinates
transverse to the geodesic. This form of the metric is well-suited
to the present calculation; the spin connection, which will be
important for the superstring action, turns out to have a simple
functional form, and the $AdS_5$ and $S^5$ subspaces appear nearly
symmetrically. This metric is invariant under the full $SO(4,2)
\times SO(6)$ symmetry, but only translation invariance in $t$ and
$\phi$ and the $SO(4)\times SO(4)$ symmetry of the transverse
coordinates remain manifest in this form. The translation
symmetries mean that string states have a conserved energy
$\omega$, conjugate to $t$, and a conserved (integer) angular
momentum $J$, conjugate to $\phi$. Boosting along the equatorial
geodesic is equivalent to studying states with large $J$ and the
lightcone Hamiltonian will give the (finite) allowed values for
$\omega-J$ in that limit. On the gauge theory side, the $S^5$
geometry is replaced by an $SO(6)$ $R$-symmetry group, and $J$
corresponds to the eigenvalue of an $SO(2)$ $R$-symmetry
generator. The AdS/CFT correspondence implies that string energies
in the large-$J$ limit should match operator dimensions in the
limit of large $R$-charge.

On dimensional grounds, taking the $J\to\infty$ limit on string states is
equivalent to taking the $\widehat R\to\infty$ limit of the geometry
(in properly chosen coordinates). The coordinate redefinitions
\begin{eqnarray}
\label{rescale}
    t \rightarrow x^+
\qquad
    \phi \rightarrow x^+ + \frac{x^-}{\widehat R^2}
\qquad
    z_k \rightarrow \frac{z_k}{\widehat R}
\qquad
    y_{k'} \rightarrow \frac{y_{k'}}{\widehat R}
\end{eqnarray}
make it possible to take a smooth $\widehat R\to\infty$ limit. (The
lightcone coordinates $x^\pm$ are a bit unusual, but have been chosen
for future convenience in quantizing the worldsheet Hamiltonian).
Expressing the metric (\ref{metric}) in these new coordinates, we obtain the
following expansion in powers of $1/\widehat R^2$:
\begin{eqnarray}
\label{expndmet}
ds^2 & \approx &
2\,{dx^+}{dx^-} + {dz}^2 + {dy }^2  -
        \left( {z }^2 + {y }^2 \right) ({dx}^+)^2 + \nonumber \\
& &     \frac{1}{\widehat R^2} \left[- 2 y^2 dx^- dx^+
+\frac{1}{2} \left( {y }^4 - {z }^4 \right) (dx^+)^2 + \left(d x^-\right)^2
    + \frac{1}{2}z^2 dz^2 - \frac{1}{2} y^2 dy^2 \right]
    \nonumber \\ & &   + {\cal O}(1/\widehat R^4)\ .
\end{eqnarray}
The leading $\widehat R$-independent part is the Penrose limit, or pp-wave
geometry: it describes the geometry seen by the infinitely boosted string.
For future reference, we define this limiting metric as
\begin{eqnarray}
\label{ppmetric}
ds^2_{pp} =
2\,{dx^+}{dx^-} + {dz }^2 + {dy }^2  -
        \left( {z }^2 + {y }^2 \right) ({dx}^+)^2\ .
\end{eqnarray}
The $x^+$ coordinate is dimensionless, $x^-$ has dimensions of
length squared, and the transverse coordinates now have dimensions of length.

In light-cone gauge quantization of the string dynamics, one
identifies world-sheet time $\tau$ with the $x^+$ coordinate, so
that the world-sheet Hamiltonian corresponds to the conjugate
space-time momentum $P_+=\omega-J$. Additionally, one sets the
world-sheet momentum density $p_{-} =1$ so that the other
conserved quantity carried by the string, $P_-=J/\widehat R^2$, is
encoded in the length of the $\sigma$ interval. Once $x^\pm$ are
eliminated, the quadratic dependence of $ds^2_{pp}$ on the
remaining eight transverse bosonic coordinates leads to a
quadratic (and hence soluble) bosonic lightcone Hamiltonian $P_+$.
Things are less simple when $1/\widehat R^{2}$ corrections to the
metric are taken into account: they add quartic interactions to
the lightcone Hamiltonian and lead to non-trivial shifts in the
spectrum of the string. This phenomenon, generalized to the
superstring, will be the primary subject of the rest of the paper.

While it is clear how the Penrose limit can bring the bosonic
dynamics of the string under perturbative control, the RR field
strength survives this limit and causes problems for quantizing
the superstring. The GS action is the only practical
approach to quantizing the superstring in RR backgrounds, and we
must construct this action for the IIB superstring in the $AdS_5
\times S^5$ background \cite{MetTseyt}, pass to lightcone
gauge and then take the Penrose limit. The latter step reduces the
otherwise extremely complicated action to a worldsheet theory of
free, equally massive transverse bosons and fermions
\cite{Metsaev:2001bj}. For reference, we give a concise summary of
the construction and properties of the lightcone Hamiltonian
$H^{GS}_{pp}$ that describes the superstring in this limit. This
will be a helpful preliminary to our principal goal of evaluating
the corrections to the Penrose limit of the GS action.

Gauge fixing eliminates both lightcone coordinates $x^\pm$,
leaving eight transverse coordinates $x^I$ as bosonic dynamical
variables. Type IIB supergravity has two ten-dimensional
supersymmetries that are described by two sixteen-component
Majorana--Weyl spinors of the same ten-dimensional chirality. The
GS superstring action contains just such a set of
spinors (so that the desired spacetime supersymmetry comes out
`naturally'). In the course of lightcone gauge fixing, half of
these fermi fields are set to zero, leaving behind a complex
eight-component worldsheet fermion $\psi$. This field is further
subject to the condition that it transform in an ${\bf 8}_s$
representation under $SO(8)$ rotations of the transverse
coordinates (while the bosons of course transform as an ${\bf
8}_v$). In a sixteen-component notation
the restriction of the world-sheet fermions to the ${\bf
8}_s$ representation is implemented by the condition $\gamma^9
\psi =+\psi$ where $\gamma^9=\gamma^1\cdots\gamma^8$ and the
$\gamma^A$ are eight real, symmetric gamma matrices satisfying a
Clifford algebra $\{\gamma^A,\gamma^B\}=2\delta^{AB}$. Another
quantity, which proves to be important in what follows, is
$\Pi\equiv\gamma^1\gamma^2\gamma^3\gamma^4.$ One could also define
$\tilde\Pi = \gamma^5\gamma^6\gamma^7\gamma^8$, but $\Pi\psi
=\tilde\Pi\psi$ for an ${\bf 8}_s$ spinor.

In the Penrose limit, the lightcone GS superstring action takes the form
\begin{equation}
\label{ppwavact}
S_{pp} = \frac{1}{2\pi\alpha^{\prime}} \int d\tau
\int_0^{2\pi\alpha^{\prime} P_-} d \sigma ({\cal L}_B+{\cal L}_F)\ ,
\qquad {\rm where}
\end{equation}
\begin{eqnarray}
\label{LBandLF}
{\cal L}_B = \frac{1}{2} \left[ ( \dot x^A)^2
-  (x^{\prime A})^2 - (x^A)^2\right]~,
\qquad
{\cal L}_F = i \psi^{\dagger} \dot\psi + \psi^{\dagger} \Pi
\psi +\frac{i}{2} ( \psi \psi^{\prime} + \psi^{\dagger}
\psi^{\prime\dagger} ).
\end{eqnarray}
The fermion mass term $\psi^{\dagger} \Pi \psi$ arises from the
coupling to the background RR 5-form field strength, and matches
the bosonic mass term (as required by supersymmetry). It is
important that the quantization procedure preserve supersymmetry.
However, as is typical in lightcone quantization, some of the
conserved generators are linearly realized on the $x^A$ and
$\psi^\alpha$, and others have a more complicated non-linear
realization.

The equation of motion of the transverse string coordinates is
\begin{equation}
\label{ppeqn} \ddot x^{A}- x^{\prime\, \prime A} +  x^A=0\ .
\end{equation}
The requirement that $x^A$ be periodic in the worldsheet
coordinate $\sigma$ (with period $2\pi\ap P_-$) leads to the mode
expansion
\begin{equation}
\label{modexpn}
x^A(\sigma, \tau) = \sum_{n= -\infty}^{\infty}x_n^A (\tau)
e^{-ik_n\sigma}~, \qquad  k_n = \frac{n}{\alpha^{\prime} P_-}
= \frac{n \widehat R^2}{\alpha^{\prime} J}~.
\end{equation}
The canonical momentum $p^A$ also has a mode expansion, related to that
of $x^A$ by the free-field equation $p^A=\dot x^A$. The coefficient
functions are most conveniently expressed in
terms of harmonic oscillator raising and lowering operators:
\begin{eqnarray}
\label{candaops}
x_n^A(\tau) =\frac{i}{\sqrt{2\omega_n P_-}}
(a_n^A e^{-i\omega_n\tau}-a_{-n}^{A\dagger} e^{i\omega_n\tau})\ ,\quad
p_n^A(\tau) = \sqrt{ \frac{\omega_n}{2 P_-}}
(a_n^A e^{-i\omega_n\tau}+a_{-n}^{A\dagger} e^{i\omega_n\tau})~.
\end{eqnarray}
The harmonic oscillator frequencies are determined by the equation
of motion (\ref{ppeqn}) to be
\begin{equation}
\label{oscens}
\omega_n=\sqrt{1+k_n^2}=\sqrt{1+({n\Rhat^2}/{\ap J})^2}=
    \sqrt{1+({g^2_{YM}N_c n^2}/{J^2})}~,
\end{equation}
where the mode index $n$ runs from $-\infty$ to $+\infty$.
(Because of the mass term, there is no separation into
right-movers and left-movers). The canonical commutation relations
are satisfied by imposing the usual creation and annihilation
operator algebra:
\begin{equation}
\label{candaopalg}
\left[a_m^A,
a_n^{B\dagger}\right] =
    \delta_{mn}\delta^{AB}~
\Rightarrow ~ \left[ x^A(\sigma),p^B(\sigma^\prime)\right]=
     i 2\pi\ap\delta(\sigma-\sigma^\prime)\delta^{AB}~.
\end{equation}

The fermion equation of motion is
\begin{equation}
i(\dot\psi +\psi^{\prime\dagger} ) + \Pi \psi =0\ .
\end{equation}
The expansion of $\psi$ in terms of creation and annihilation operators
is achieved by expanding the field in worldsheet momentum eigenstates
\begin{equation}
\psi(\sigma, \tau) = \sum_{n= -\infty}^{\infty}\psi_{n} (\tau) 
e^{-ik_n\sigma}~,
\end{equation}
which are further expanded in terms of convenient positive and negative
frequency solutions of the fermion equation of motion:
\begin{equation}
\label{fmodexp}
\psi_n(\tau) = \frac{1}{\sqrt{4P_-\omega_n}}
    (e^{-i\omega_n\tau}(\Pi+\omega_n-k_n) b_n
     +e^{i\omega_n\tau}(1-(\omega_n-k_n)\Pi) b^\dagger_n )~.
\end{equation}
The frequencies and momenta in this expansion are equivalent to
those of the bosonic coordinates. In order to reproduce the
anticommutation relations
\begin{equation}
\{\psi(\tau,\sigma),\psi^\dagger(\tau,\sigma^\prime)\}
    = 2\pi\ap\delta(\sigma-\sigma^\prime)~,
\end{equation}
we impose the standard oscillator algebra
\begin{equation}
\{ b_m^\alpha , b_n^{\beta\dagger} \} =
\frac{1}{2} (1 + \gamma_9)^{\alpha\beta} \delta_{m,n}~.
\end{equation}
The spinor fields $\psi$ carry sixteen components, but the ${\bf 8}_s$
projection reduces this to eight anticommuting oscillators,
exactly matching the eight transverse oscillators in the bosonic
sector. The final expression for the light-cone Hamiltonian is
\begin{equation}\label{ppham}
 H^{GS}_{pp} = \sum_{n= -\infty}^{+\infty}\omega_n
\left( \sum_A (a_n^A)^\dagger a_n^A +
\sum_\alpha (b_n^\alpha)^\dagger b_n^\alpha \right)\ .
\end{equation}
The harmonic oscillator zero-point energies nicely cancel between
bosons and fermions for each mode $n$. The frequencies $\omega_n$
depend on the single parameter
\begin{equation}
\lambda^{\prime}= g^2_{YM}N_c/J^2 \qquad
    \omega_n=\sqrt{1+\lambda^\prime n^2},
\end{equation}
so that one can take $J$
{\sl and} $g^2_{YM}N_c$ to be simultaneously large while keeping
$\lambda^{\prime}$ fixed. If $\lambda^{\prime}$ is kept fixed and
small, $\omega_n$ may be expanded in powers of $\lambda^{\prime}$,
suggesting that contact with perturbative Yang--Mills gauge theory
is possible. 

The spectrum is generated by $8+8$ transverse oscillators acting
on ground states labeled by an $SO(2)$ angular momentum taking
integer values $-\infty<J<\infty$ (note that the oscillators
themselves carry zero $SO(2)$ charge). Any combination of
oscillators may be applied to a ground state, subject to the
constraint that the sum of the oscillator mode numbers must vanish
(this is the level-matching constraint, the only constraint not
eliminated by lightcone gauge-fixing). The energies of these
states are the sum of the individual oscillator energies
(\ref{oscens}), and the spectrum is very degenerate.\footnote{Note
that the $n=0$ oscillators raise and lower the string energy by a
protected amount $\delta P_+=1$, independent of the variable
parameters. These oscillators play a special role, enlarging the
degeneracy of the string states in a crucial way, and we will call
them `zero-modes' for short.} For example, the 256 states of the
form $A^\dagger_n B^\dagger_{-n} \vert J\rangle$ for a given mode
number $n$ (where $A^\dagger$ and $B^\dagger$ each can be any of
the 8+8 bosonic and fermionic oscillators) all have the energy
\begin{equation} \label{twoimpen}
P_+= \omega-J = 2\sqrt{1+({
g^2_{YM}N_c n^2}/{J^2})}
    \sim {2+({ g^2_{YM}N_c n^2}/{J^2})+\ldots}\, .
\end{equation}
In the weak coupling limit ($\lambda^\prime\to 0$) the degeneracy is
even larger because the dependence on the oscillator mode number $n$
goes away! This actually makes sense from the dual gauge theory point
of view where $P_+\to D-R$ ($D$ is the dimension and $R$ is the $R$-charge
carried by gauge-invariant operators of large $R$); at zero coupling,
operators have integer dimensions and the number of operators with
$D-R=2$, for example, grows with $R$, providing a basis on which string
multiplicities are reproduced.  Even more remarkably, BMN were able to show
\cite{Berenstein:2002jq} that subleading terms in a $\lambda^{\prime}$
expansion of the string energies match the first perturbative corrections
to the gauge theory operator dimensions in the large $R$-charge limit. We will
review the details of this agreement in the next section.  

More generally, we expect exact string energies in the
$AdS_5\times S^5$ background to have a joint expansion in the
parameters $\lambda^{\prime}$, defined above, and $1/J$. 
We also expect the degeneracies found in the $J\to\infty$ limit (for fixed
$\lambda^\prime$) to be lifted by interaction terms that arise in
the worldsheet Hamiltonian describing string physics at large but
finite $J$. Large degeneracies must nevertheless remain in order
for the spectrum to be consistent with the $PSU(2,2\vert 4)$
global supergroup that should characterize the exact string
dynamics. The specific pattern of degeneracies should also match
that of operator dimensions in the ${\cal N}=4$ super Yang--Mills
theory. Since the dimensions must be organized by the
$PSU(2,2\vert 4)$ superconformal symmetry of the gauge theory,
consistency is at least possible, if not guaranteed. In the rest of this
paper we will explore this question.  We will first summarize the
information about gauge theory operator anomalous dimensions that
is needed to address these issues. We will then describe our
evaluation of the interaction terms that must be added to
$H^{GS}_{pp}$ to accommodate corrections to superstring worldsheet
physics in the large-$J$ limit. Finally, we will report the
results of a first-order degenerate perturbation theory treatment
of these corrections to the string worldsheet Hamiltonian, and
show that they precisely match the relevant gauge theory
expectations.
\section{Gauge Theory, Group Theory, Dimension Expansion}

Before tackling the calculation of the energy spectrum of the
string, we first present some information on dimensions of gauge
theory operators that will be needed to test the predictions of
this duality. Most of what we will say in this section can be 
found in the literature in one form or another, especially
in work by Beisert \cite{BeisertSUSY}. However, in order to organize 
things in the most suitable way for our subsequent comparison
with string theory results (and to make a few points that don't
seem to have been made elsewhere), we found it useful, at least
for ourselves, to rederive and restate mostly known results. 
We focus on the noninteracting string ($g_s\to 0$)
which, as explained earlier, is dual to the gauge theory in the
large-$N_c$ limit (the Yang-Mills genus-counting parameter is $g_2 = J^2/N_c$  
\cite{GenusCounting}). In the large-$N_c$ limit, the operators of
interest are single-gauge-trace monomials of fields of ${\cal
N}=4$ SUSY Yang--Mills theory.\footnote{Multiple trace operators
appear when we go beyond the large-$N_c$ limit; 
they mix with single-trace operators when non-planar diagrams
are included.} In the string theory, we look at states of large $J$ 
but finite $P_+=\omega-J$.

In the gauge theory we classify operators by $R$-charge (in some
$SO(2)$ subgroup of the $SU(4)$ $R$-symmetry group) and dimension
$D$. To match the kinematic limit of the string states, we look
for operators with dimension $D$ and $R$ both large but with
$\Delta=D-R$ finite. The dimension $D$ approaches the naive engineering
dimension, which we will denote by $K$, in the limit $\lambda' \to 0$.
Thus, the limit of interest is $K,R\to\infty$ with the integer difference
$\Delta_0 = K - R$ held fixed. Thus $\Delta$ will be the sum of
the integer $\Delta_0$ and the anomalous dimension $D-K$ of the operator
(which we assume to have a finite limit). Our problem is to
identify an appropriate basis of gauge operator monomials which are
mixed by the dimension operator, calculate the anomalous dimension
matrix to some order in perturbation theory, and then diagonalize
that matrix to get the allowed values of $\Delta$. The question
is whether they match the string theory spectrum of $P_+$.

The component fields available to us in ${\cal N}=4$ SYM are a
gauge field, a set of gluinos and a set of scalars, all in the
adjoint of the gauge group. The theory has an exact global $SU(4)$
$R$-symmetry, under which the gluinos transform as a $\mathbf{4}$
and $\mathbf{\bar 4}$ and the scalars as a $\mathbf{6}$. Since the
dimension matrix commutes with the full $R$-symmetry group, it is
helpful to classify operators according to their $SU(4)$
representation. Irreducible tensor representations of $SU(4)$ are
indexed by Young diagrams describing their symmetries under
permutations of the tensor indices. Such diagrams contain up to
three rows of boxes with non-increasing numbers of boxes per row
and are denoted by a set of three integers $(n_1,n_2,n_3)$ giving
the differences in length of successive rows. The total number of
boxes in the diagram is the total number of $SU(4)$ indices in the
tensor. The boxes are filled in with tensor indices in some
canonical order and the representations are antisymmetric under
the exchange of any pair of indices in the same column. More
specifically, the scalars are in the 6-dimensional $(0,1,0)$
representation of $SU(4)$, the gluinos are 2-component Weyl
spacetime spinors in the 4-dimensional fundamental $(1,0,0)$ plus an
adjoint field in the 4-dimensional anti-fundamental $(0,0,1)$:
\begin{equation}
{\rm Scalars:}~~\suphi \qquad {\rm Gluinos:}~~
\suchi_{~a},~\suchib_{~\dot a}~.
\end{equation}
The $a$ (resp.~$\dot a$) indices on the gluinos indicate that they
transform in the $(\mathbf{2,1})$ (resp. $(\mathbf{1,2})$)
representations of the $SL(2,C)$ covering group of the spacetime
Lorentz group. All of these fields, as well as the $SU(4)$-singlet
gauge field, are adjoint matrices in the gauge group algebra. The
Young diagram superscript is a shorthand for indicating the
$SU(4)$ tensor character of the fields (viz.~$\phi$ is a rank-two
antisymmetric tensor, $\chi_a$ is a rank-one tensor and so on).

We want to use these fields to construct gauge-singlet composite operators.
As mentioned before, we work in the leading large-$N_c$ limit and need
only consider monomials involving a single gauge trace. For the moment,
we limit our attention to operators that are spacetime scalars. The $SO(2)$
scalar $R$-charge that will eventually be taken to infinity (to match the
$J\to\infty$ limit of the string spectrum) is defined by the decomposition
$SU(4)\supset SU(2)\times SU(2)\times U(1)_R$ (the same thing as
$SO(6)\supset SO(4)\times SO(2)$). The scalar $R$-charge of the various
components of the gauge theory fields is assessed by distributing indices
in the boxes of the Young diagram superscripts, subject to the rule of
column antisymmetry and assigning $R=\frac{1}{2}(-\frac{1}{2})$ to
$SU(4)$ indices $1,2~(3,4)$ respectively. The result is as follows:
\begin{eqnarray}
R=1:    ~\phi^{\,\Yboxdim5pt\tiny\young(1,2)}~ (Z), \qquad
R=0:    ~\phi^{\,\Yboxdim5pt\tiny\young(1,3)},
    \phi^{\,\Yboxdim5pt\tiny\young(1,4)},
    \phi^{\,\Yboxdim5pt\tiny\young(2,3)},
    \phi^{\,\Yboxdim5pt\tiny\young(2,4)}~ (\phi^A), \qquad
R=-1:   ~\phi^{\,\Yboxdim5pt\tiny\young(3,4)}~ (\bar Z), \nonumber\\
R=1/2:  ~\chi^{\,\Yboxdim5pt\tiny\young(1)},
    \chi^{\,\Yboxdim5pt\tiny\young(2)},
    \bar\chi^{\,\Yboxdim5pt\tiny\young(1,2,3)},
    \bar\chi^{\,\Yboxdim5pt\tiny\young(1,2,4)}, \qquad
R=-1/2: ~\chi^{\,\Yboxdim5pt\tiny\young(3)},
    \chi^{\,\Yboxdim5pt\tiny\young(4)},
    \bar\chi^{\,\Yboxdim5pt\tiny\young(1,3,4)},
    \bar\chi^{\,\Yboxdim5pt\tiny\young(2,3,4)}~.
\end{eqnarray}
We have introduced an alternate notation for the scalars (to be
used later) ($Z,\bar Z,\phi^A, A=1,..,4$) that emphasizes their
$SO(4)$ content. As discussed earlier, we need a basis of
operators with large naive dimension $K$, large scalar $R$-charge and
fixed $\Delta_0 =K-R$. BMN showed that, in this limit, such
operators correspond to string states created by a fixed finite
number ($\Delta_0$) of string oscillators acting on the pp-wave
ground state of angular momentum $R$. Operators with $\Delta_0 =
0$ are BPS, and their dimensions are protected by supersymmetry.
In what follows, we will, for simplicity, restrict the discussion
to $\Delta_0=2$ operators, corresponding to string states created
by two oscillators acting on the vacuum (so-called `two-impurity'
states). The list of {\sl all} single-trace spacetime scalar
operators of naive dimension $K$ which can have $\Delta_0\le 2$ is
as follows:
\begin{eqnarray}
\label{optypes}
\tr\big((\suphi)^K\big),\qquad& \qquad (R_{\rm max} = K ) \nonumber\\
\tr\big((\suchi\sigma_2\suchi)(\suphi)^{K-3}\big),
~\tr\big((\suchi\suphi\sigma_2\suchi)(\suphi)^{K-4}\big),& ~\ldots
    \qquad (R_{\rm max} = K-2) \nonumber\\
\tr\big((\suchib\sigma_2\suchib)(\suphi)^{K-3}\big),
~\tr\big((\suchib\suphi\sigma_2\suchib)(\suphi)^{K-4}\big),&~ \ldots
    \qquad (R_{\rm max} = K-2) \nonumber\\
\tr\big(\nabla_\mu\suphi\nabla^\mu\suphi(\suphi)^{K-4}\big),
    \qquad& \qquad (R_{\rm max} = K-2 )\ .
\end{eqnarray}
The fields inside the operators are $SU(N_c)$ adjoint matrices and
the trace is taken over gauge indices; spacetime spinor indices on
the $\chi$ are contracted to produce a spacetime scalar (note that
a product of a $\suchi$ and a $\suchib$ cannot make a scalar
because they transform under inequivalent irreps of spacetime
$SL(2,C)$); $\nabla$ is the spacetime gauge-covariant derivative.
There are multiple versions of operators involving gluinos and
spacetime derivatives arising from the different ways that scalars
may be distributed among them (and the cyclic symmetry of the
gauge trace reduces the number of independent operators one can
construct). These operators provide a basis for a reducible
representation of the global $SU(4)$ $R$-symmetry group. Since the
anomalous dimension operator commutes with this $SU(4)$, it will
have no matrix elements between different $SU(4)$ irreps, and our
first task is to find linear combinations of the above operators
that provide a basis for these irreps (and find the multiplicities
of inequivalent occurrences of the same irrep). The group theory
analysis helps us obtain precise control of the subleading
corrections in $1/K$ to the structure of the operators and their
anomalous dimensions.

For the bosonic operators with no derivatives, we have a reducible
$SU(4)$ tensor of rank $2K$ which we must decompose into
irreducible $SU(4)$ tensors of rank $2K$. These irreps are
symbolized by Young diagrams with $2K$ boxes; the main problem is
to determine the multiplicity with which each such diagram
appears. The standard algorithm for projecting a reducible
character onto irreducible characters \cite{Boerner} cannot be
implemented because of the cyclic symmetry of single-trace
monomials. The algorithm, however, can be adapted with some effort
to the case at hand to compute the desired
multiplicities\footnote{We report here only the pertinent results
and leave the exposition of the group theory particulars to a
longer publication}. Although the total number of irreducible
tensors in the expansion grows rapidly with $K$, only a few can
have $\Delta_0=K-R=0,2$ and we report only the multiplicities of
that limited set of irreps. The results are slightly different for
odd and even $K$, but we will eventually see that this even/odd
difference is harmless.  For $K$ odd we have
\begin{eqnarray}
\label{phi_irrepodd}
\Yvcentermath1
\tr\big({\suphi}^{~K}~\big) \to~
    1\times{\underbrace{\tiny\yng(7,7)}_{ K}} ~\oplus ~~
\Yvcentermath1
\left(\frac{K-1}{2}\right)\times{\underbrace{\tiny\yng(6,4)}_{K-1}} ~\oplus ~~
\left(\frac{K-1}{2}\right)\times{\underbrace{\tiny\yng(5,5)}_{K-2}}~\oplus 
\nonumber\\
\Yvcentermath1
\left(\frac{K-1}{2}\right)\times{\underbrace{\tiny\yng(6,6,2)}_{K-1}} 
~\oplus ~~
\left(\frac{K-3}{2}\right)\times{\underbrace{\tiny\yng(7,5,2)}_K}
~\oplus~~ \dots\ ,
\end{eqnarray}
while for $K$ even we have
\begin{eqnarray}
\label{phi_irrep}
\Yvcentermath1
\tr\big({\suphi}^{~K}~\big) \to~
    1\times{\underbrace{\tiny\yng(7,7)}_{ K}} ~\oplus ~~
\Yvcentermath1
\left(\frac{K-2}{2}\right)\times{\underbrace{\tiny\yng(6,4)}_{K-1}} ~\oplus
~~ \left(\frac{K}{2}\right)\times{\underbrace{\tiny\yng(5,5)}_{K-2}}~\oplus
\nonumber\\ \Yvcentermath1
\left(\frac{K-2}{2}\right)\times{\underbrace{\tiny\yng(6,6,2)}_{K-1}} ~\oplus
~~ \left(\frac{K-2}{2}\right)\times{\underbrace{\tiny\yng(7,5,2)}_K}
~\oplus~\dots~~
\end{eqnarray}
These irrep expansions could equally well have been done using the
bosonic R-symmetry group $SO(6)$: this is what is done, with the same
results, in \cite{BeisertSUSY}.
The irreps with larger minimal values of $\Delta_0=K-R$ have
multiplicities that grow as higher powers of $K$. This is very
significant for the eventual string theory interpretation of the
anomalous dimensions, but we will not expand on this point here.

Other spacetime scalar operators that can have $\Delta_0=K-R=2$
are the `bifermions', or products of two gluinos and $K-3$
scalars. Including only the irreps that can actually have
$\Delta_0=2$, their expansions are as follows:
\begin{eqnarray}
\label{chi_irrep} \Yvcentermath1
\tr\big(\suchi~\sigma_2~\suchi~ {(\suphi)}^{ K-3}\big) \to~
    1\times{\underbrace{\tiny\yng(5,5)}_{K-2}} ~\oplus ~~
1\times{\underbrace{\tiny\yng(6,4)}_{K-1}} ~\oplus ~\ldots
\end{eqnarray}
\begin{eqnarray}
\label{chib_irrep}
\Yvcentermath1
{\tr}\big(\suchib~\sigma_2~\suchib~
{(\suphi)}^{ K-3}\big) \to~
    1\times{\underbrace{\tiny\yng(6,6,2)}_{K-1}} ~\oplus ~~
1\times{\underbrace{\tiny\yng(5,5)}_{K-2}} ~\oplus ~\ldots
\end{eqnarray}
There are identical expansions for operators arising from different
placements of the fermions with respect to the bosons. Because of cyclicity
of the gauge trace and the fermi statistics of the gluino fields, these
operators are not all independent. The counting of independent operators
depends, once again, on whether $K$ is even or odd. Using an obvious
shorthand notation, the multiplicities of bifermion irreps are as
follows for $K$ odd:
\begin{eqnarray}
\label{gluino_irrep_odd}
\Yvcentermath1
\tr\big(\suchi~\sigma_2~\suchi~
{(\suphi)}^{ K-3}\big) \to~
\Yvcentermath1
\left(\frac{K-3}{2}\right)\times{\underbrace{\tiny\yng(5,5)}_{K-2}} ~\oplus ~~
\left(\frac{K-1}{2}\right)\times{\underbrace{\tiny\yng(6,4)}_{K-1}} 
~\oplus~\ldots ~~\\
\Yvcentermath1
{\tr}\big(\suchib~\sigma_2~\suchib~
{(\suphi)}^{ K-3}\big) \to~
\left(\frac{K-3}{2}\right)\times{\underbrace{\tiny\yng(5,5)}_{K-2}} ~\oplus ~~
\left(\frac{K-1}{2}\right)\times{\underbrace{\tiny\yng(6,6,2)}_{K-1}}
    ~\oplus~\ldots ~~
\end{eqnarray}
The results for $K$ even are, once again, slightly different:
\begin{eqnarray}
\label{gluino_irrep_even}
\Yvcentermath1
\tr\big(\suchi~\sigma_2~\suchi~
{(\suphi)}^{ K-3}\big) \to~
\Yvcentermath1
\left(\frac{K-2}{2}\right)\times{\underbrace{\tiny\yng(5,5)}_{K-2}} ~\oplus ~~
\left(\frac{K-2}{2}\right)\times{\underbrace{\tiny\yng(6,4)}_{K-1}} 
~\oplus ~\ldots\\
\Yvcentermath1
{\tr}\big(\suchib~\sigma_2~\suchib~
{(\suphi)}^{ K-3}\big) \to~
\left(\frac{K-2}{2}\right)\times{\underbrace{\tiny\yng(5,5)}_{K-2}} ~\oplus ~~
\left(\frac{K-2}{2}\right)\times{\underbrace{\tiny\yng(6,6,2)}_{K-1}} 
~\oplus ~\ldots
\end{eqnarray}

The point of all this is that the dimension operator can only have
matrix elements between operators belonging to the same $SU(4)$
irrep. There is a unique irrep, $(0,K,0)$ (i.e., two rows of $K$
boxes), which contains `top' states with dimension equal to
$R$-charge (or $\Delta_0 = \Delta =0$). The latter are known to be
BPS states and get no correction to their dimension. Thus the
dimension of the whole irrep, including all its components with
$\Delta_0 >0$, is unmodified by interactions. The other irreps
displayed above have multiplicities that grow roughly as $K/2$ for
large $K$. The irreps we have not displayed have higher values of
$\Delta_0$ {\sl and} multiplicities that grow as higher powers of
$K$. The dimension operator will, in general, have matrix elements
between all the operators belonging to a given representation. We
therefore have to diagonalize a matrix of size roughly $K/2\times
K/2$ and will find $O(K/2)$ eigenvalues. The key question will
then be the evolution of the spectrum as $K\to\infty$. From the
work of BMN, we expect to find a spectrum that can be interpreted,
at large $K=R+2$ and fixed $\Delta_0=K-R = 2$, as due to the action of two
string modes on a string ground state of angular momentum $J=R$. Our
goal is to evaluate and compare the $1/R$ corrections on both
sides of this correspondence. One benefit of the group theory
analysis is immediately apparent: the irrep $(2,K-4,2)$ appears
{\em  only} in the reduction of the purely bosonic operator. For
this irrep, the anomalous dimension matrix must act purely within
the space of bosonic operators, a welcome simplification. By
contrast, the irrep $(0,K-3,2)$ appears both in the purely bosonic
operators and in one of the two-fermion operators (with the same
multiplicity in both cases). Thus, there can be matrix elements of
the dimension operator between boson and fermion states and the
diagonalization problem will be more complicated. In fact, the
results of the diagonalization will test the fermionic structure
of the string Hamiltonian, which makes this a particularly
important test to carry out.

Having calculated the multiplicity of specific irreps, we turn to the
perturbative diagonalization of the dimension operator. A simple approach
begins with the two-point function between elements of the operator basis
$\left\{O_a(x)\right\}$, calculated to first non-trivial order in
perturbation theory. The typical result is
\begin{equation}
\label{opprodexp} \langle O_a(x)O_b(0)\rangle \sim
(x)^{-2d_0}(\delta_{ab}+\ln(x^2)d_1^{ab})\ ,
\end{equation}
where $d_0$ is the naive dimension. The leading Kronecker
$\delta_{ab}$ implies that the operator basis is orthonormal in
the free theory (in the large-$N_c$ limit, this is enforced by
multiplying the operator basis by a common overall normalization
constant). The anomalous dimensions are then the eigenvalues of
the mixing matrix $d_1^{ab}$, and the eigenoperators of definite
dimension are linear combinations of basis operators defined by
the eigenvectors.  One should be careful to pick out conformal
primary operators, but this subtlety is not too troublesome for
one-loop perturbative calculations.

Group theory tells us that the dimension operator $D$
block-diagonalizes under the different $SU(4)$ irreps, and it is
not too hard to show in concrete detail how it works in the purely
bosonic sector. Consider a basis of $K-1$ bosonic operator
monomials of dimension $K$ and $\Delta_0=K-R=2$:
\begin{eqnarray}
\label{singltrbasis}
\lbrace O^{AB}_{K,1},\ldots,O^{AB}_{K,K-1}
\rbrace =
    \lbrace \tr(ABZ^{K-2}),~\tr(AZBZ^{K-3}),\ldots,~\nonumber\\
        \tr(AZ^{K-3}BZ),~\tr(AZ^{K-2}B) \rbrace~,
\end{eqnarray}
where $Z$ stands for $\labphi{1}{2}$ and has $R=1$, while
$A,B$ stand for any of the four $\phi^A$ ($A=1,\ldots,4$) and have
$R=0$. The overall constant needed to orthonormalize this basis
(in the large-$N_c$ limit) is easy to compute, but not needed for
present purposes. In the $SO(2)\times SO(4)$ decomposition of
$SU(4)$, $A,B$ are $SO(4)$ vectors so that the operators of this
basis are rank-two $SO(4)$ tensors. We won't give the detailed
argument here, but it can be shown that the the members of this
basis can be assigned to $SU(4)$ irreps by splitting them into
irreducible rank-two $SO(4)$ tensors. In particular, the symmetric
traceless tensor belongs to the $(2,K-4,2)$ irrep of $SU(4)$, the
antisymmetric tensor belongs to the pair $(0,K-3,2)+(2,K-3,0)$,
and the $SO(4)$ trace (when completed to a full $SO(6)$ trace)
belongs to the $(0,K-2,0)$ irrep. In what follows, we refer to
these three classes of operator as $\overline T_K^{(+)}$,
$\overline T_K^{(-)}$ and $\overline T_K^{(0)}$, respectively. If
we take $A\ne B$, the trace part drops out and the $\overline
T_K^{(\pm)}$ operators are isolated by symmetrizing and
antisymmetrizing on $A,B$.

A simple extension of the BMN argument can be used to give the
$O(g^2_{YM}N_c)$ action of the anomalous dimension operator on the
basis (\ref{singltrbasis}), correct to all orders in $1/K$. In the
leading large-$N_c$ limit and leading order in
$g^2_{YM}$, the gauge theory interaction term
$\tr([\phi^a,\phi^b][\phi^a,\phi^b])$ has a very simple action on
single-trace monomials in the $\phi$'s: it produces a sum of
interchanges of all nearest-neighbors in the trace. Diagrams that
lead to exchanges at greater distances are non-planar and
suppressed by powers of $1/N_c$. For the restricted case $A\ne B$,
the leading action of the anomalous dimension on the $K-1$ bosonic
monomials of (\ref{singltrbasis}) has the following detailed
structure:
\begin{eqnarray}\label{bosopermute}
(ABZ^{K-2})\rightarrow
(BAZ^{K-2})+2(AZBZ^{K-3})+(K-3)(ABZ^{K-2})\nonumber\\
(AZBZ^{K-3})\rightarrow
2(ABZ^{K-2})+2(AZ^2BZ^{K-4})+(K-4)(AZBZ^{K-3})\nonumber\\
    \ldots\ldots\qquad\qquad \nonumber\\
(AZ^{K-2}B)\rightarrow
2(AZ^{K-3}BZ)+(K-3)(BAZ^{K-2})+(ABZ^{K-2})~,
\end{eqnarray}
(omitting the
overall factor coming from the details of the Feynman diagram).
The action on the trace parts when $A=B$ is more complicated, and we
will omit the detailed argument for that case. In an obvious
matrix notation, we have
\begin{equation}
\label{bigmat}
\bigl[~{\rm
Anom~~Dim}~\bigr]_{(K-1)\times(K-1)}~\sim~
\begin{pmatrix}
    K-3&2&0&\ldots&1\cr
    2&K-4&2&\ldots&0\cr\cr
    0&\ldots&2&K-4&2\cr
    1&\ldots&0&2&K-3\cr
\end{pmatrix}~.
\end{equation}

The logic of renormalization theory allows for a subtraction
on the diagonal of this matrix, and in fact one is needed. The
vector $\vec X_0=(1,\ldots,1)$, corresponding to the operator in
which all operators in (\ref{singltrbasis}) are summed over with
equal weight, is an eigenvector with eigenvalue $K$. This
particular operator actually belongs to the special representation
$(0,K,0)$, whose anomalous dimensions must vanish because it
contains the chiral primary BPS operator $\tr(Z^K)$ (whose
dimension is equal to the $R$-charge). To properly normalize
(\ref{bigmat}) and ensure that this eigenvector has eigenvalue
zero, we subtract $K$ times the unit matrix and drop the zero
eigenvector of the anomalous dimension matrix on the grounds that
it belongs to the `uninteresting' $(0,K,0)$ representation. The
anomalous dimensions we seek are therefore the non-zero
eigenvalues of the matrix
\begin{equation}
\label{renormbigmat}
\bigl[~{\rm
Anom~~Dim}~\bigr]_{(K-1)\times(K-1)}~\sim~
\begin{pmatrix}
        -3&+2&0&\ldots&1\cr
        +2&-4&+2&\ldots&0\cr\cr
        0&\ldots&+2&-4&+2\cr
        +1&\ldots&0&+2&-3\cr
\end{pmatrix}~.
\end{equation}

This looks very much like the lattice Laplacian for a particle
hopping from site to site on a periodic lattice. The special
structure of the first and last rows assigns an extra energy to
the particle when it hops past the origin. This breaks strict
lattice translation invariance but makes sense as a picture of the
dynamics involving two-impurity states: the impurities propagate
freely when they are on different sites and have a contact
interaction when they collide. This picture has lead people to
map the problem of finding operator dimensions onto the technically
much simpler one of finding the spectrum of an equivalent
quantum-mechanical Hamiltonian \cite{QMforDim}. In one version, the map is
to a spin-chain system with integrable dynamics \cite{minahan},
suggesting that exact results for many quantities of interest
may be possible. This is an important topic, but we will not pursue 
it further in this paper.

Before diagonalizing (\ref{renormbigmat}), we note a useful
symmetry of the problem: the operator monomials in the basis
(\ref{singltrbasis}) go into each other pairwise under $A\leftrightarrow B$
and, at the same time, the vector $\vec C=(C_1,\ldots,C_{K-1})$
representing a linear combination of monomials transforms as
$C_i\to C_{K-i}$. Since (\ref{renormbigmat}) is invariant under
this transformation, its eigenvectors will be either even
($C_i=C_{K-i}$) or odd ($C_i=-C_{K-i}$) under it. Since the two
options (even or odd under $A\leftrightarrow B$) correspond to different
$SU(4)$ irreps, assessing the $SU(4)$ assignment of the different
eigenvalues will be easy. The two classes of eigenvalues and
normalized eigenvectors are as follows:
\begin{eqnarray}
\label{symspect}
\lambda^{(K+)}_{n}~=~8\sin^2\left(\frac{n\pi}{K-1}\right) \qquad
n=(0),1,2,\ldots, n_{max} = \begin{cases}(K-3)/2~~K~{\rm odd}\cr
(K-2)/2~~K~{\rm even} \end{cases}~, \nonumber\\
   C^{(K+)}_{n,i} =
\frac{2}{\sqrt{K-1}}\cos\left[\frac{2\pi n}{K-1}(i-\frac{1}{2})\right]
    \qquad i=1,\ldots,K-1~, \qquad
\end{eqnarray}
\begin{eqnarray}
\label{antisymspect}
\lambda^{(K-)}_{n}~=~8\sin^2\left(\frac{n\pi}{K}\right) \qquad n=1,2,\ldots,
n_{max} = \begin{cases}(K-1)/2~~K~{\rm odd}\cr (K-2)/2~~K~{\rm even} 
\end{cases}~,
\nonumber\\
   C^{(K-)}_{n,i} =
\frac{2}{\sqrt{K}}\sin\left[\frac{2\pi n}{K}(i)\right]
    \qquad i=1,\ldots,K-1~.\qquad
\end{eqnarray}
For the case of $\lambda^{(K+)}_n$, we indicate that $n=0$ is a possible
eigenvalue, but we must remember that it belongs to the $(0,K,0)$ irrep
when we count irrep multiplicities. The eigenoperators corresponding to
the various dimensions are constructed from the eigenvectors according to
\begin{equation}
\label{eigenop}
  \overline T^{(\pm)}_{K,n} (x) =
    \sum_{i=1}^{K-1} C^{(K\pm)}_{n,i}O^{AB}_{K,i}(x) ~.
\end{equation}
The subscript $n$ will not be displayed in the following.

To get $\Delta=D-R$, we multiply these eigenvalues by the
appropriate overall normalization factor and add the zeroth order
value $\Delta_0=2$. The results for $\overline T_K^{(+)}$
(symmetric traceless, belonging to the $(2,K-4,2)$ irrep),
$\overline T_K^{(-)}$ (antisymmetric, belonging to the
$(0,K-3,2)+(2,K-3,0)$ irreps) and $\overline T_K^{(0)}$ (trace,
belonging to the $(0,K-2,0)$ irrep) are
\begin{eqnarray} \label{firstopdim}
{\Delta}(\overline T_K^{(+)}) = 2 +
        \frac{g^2_{YM} N_c}{\pi^2} \sin^2\left(\frac{n\pi}{K-1}\right)
    \qquad n=1,2,\ldots,n_{max} =
    \begin{cases}(K-3)/2~~K~{\rm odd}\cr (K-2)/2~~K~{\rm even} \end{cases}~,
\nonumber\\
    { \Delta}(\overline T_K^{(-)}) = 2 +
        \frac{g^2_{YM} N_c}{\pi^2} \sin^2\left(\frac{n\pi}{K}\right) \qquad
    n=1,2,\ldots, n_{max}= \begin{cases}(K-1)/2~~K~{\rm odd}\cr
        (K-2)/2~~K~{\rm even}\end{cases}~,
\nonumber\\
    { \Delta}(\overline T_K^{(0)}) = 2 +
        \frac{g^2_{YM} N_c}{\pi^2} \sin^2\left(\frac{n\pi}{K+1}\right) \qquad
    n=1,2,\ldots, n_{max}= \begin{cases}(K-1)/2~~K~{\rm odd}\cr
        (K/2)~~~~~~~K~{\rm even}\end{cases}~.
\nonumber\\
\end{eqnarray}
Note that the counting of eigenvalues corresponds
exactly to the multiplicities of these irreps as reported in
(\ref{phi_irrepodd}) and (\ref{phi_irrep}). The above results on dimensions
and eigenoperators can all be found in \cite{BeisertSUSY} and, piecemeal,
in earlier discussions of the one-loop operator dimension problem.

The expressions in (\ref{firstopdim}) are the first terms in a
perturbative expansion. Since we must work in the limit of large
$g^2_{YM} N_c$, this expansion is not guaranteed to be reliable.
The string theory discussion will show that the eigenvalue index
$n$ is to be interpreted as the mode number of an excited string
oscillator. This implies a limiting procedure in which $n$ is held
fixed while $R$ and $g^2_{YM} N_c$ are taken to infinity such that
there are two controlled, small parameters, $g^2_{YM} N_c/R^2$ and
$1/R$. We will assume, as proposed by BMN, that the smallness of
$g^2_{YM} N_c/R^2$ makes perturbation theory reliable, at least
for fixed-$n$ eigenvalues (without this assumption, there is
little one can calculate on the gauge theory side). At the same
time, the smallness of $1/R$ controls the size of interaction
corrections to the Penrose limit string worldsheet Hamiltonian. If
we express the dimension formulae (\ref{firstopdim}) in terms of
$R$-charge $R$, rather than naive dimension $K$ (using $K=R+2$)
and take the limit in this way, we find
\begin{eqnarray} \label{finlopdim}
    {\Delta}(\overline T_{R+2}^{(+)})
    \rightarrow 2 +\frac{g^2_{YM} N_c}{R^2}~n^2~
            \left(1-\frac{2}{R}+O(R^{-2})\right)~,
\nonumber\\
        {\Delta}(\overline T_{R+2}^{(-)})
        \rightarrow 2 +\frac{g^2_{YM} N_c}{R^2}~n^2~
                        \left(1-\frac{4}{R}+O(R^{-2})\right)~,
\nonumber\\
        {\Delta}(\overline T_{R+2}^{(0)})
        \rightarrow 2 +\frac{g^2_{YM} N_c}{R^2}~n^2~
                        \left(1-\frac{6}{R}+O(R^{-2})\right)~.
\end{eqnarray}
To leading order in $1/R$, the dimensions of these operator
multiplets are degenerate and agree with the corresponding
expression in the Penrose limit (\ref{twoimpen}). The degeneracy
is lifted at subleading order in $1/R$, just as the Penrose limit
degeneracy of string worldsheet energies should be lifted by
string worldsheet interactions. Our goal is show that the two
approaches to the lifting of operator dimension (string energy)
degeneracy give equivalent results on each side of the duality.

The AdS/CFT interpretation of the operator dimensions displayed in
(\ref{finlopdim}) is that they are dual to the energies of string
states built out of two bosonic mode creation operators:
$(a_n^A)^\dagger (a_{-n}^B)^\dagger|R\rangle$. It is important to
note that these anomalous dimensions are valid for {\em all}
operators in the representations in question, not just those for
which $\Delta_0=K-R = 2$; this is a simple consequence of the
global $SU(4)$ $R$-symmetry. We believe that this translates on
the string theory side into the existence of exact zero-mode
oscillators $a_0^A$ which augment the $P_+$ eigenvalue of a state
by unity, independent of $g^2_{YM} N_c/J^2$ and $1/J$. This is
true in the Penrose limit, as we can infer from (\ref{oscens}),
and we expect it to continue to be true to all orders in $1/J$. If
so, the string states
\begin{equation}
(a_n^A)^\dagger (a_{-n}^B)^\dagger
    (a_0^{C_1})^\dagger\ldots (a_0^{C_s})^\dagger |J-2-s\rangle
\end{equation}
should all have the same energy and correspond to the $\Delta_0
>2$ components of the $(2,J-4,2)$ irrep (if we project onto
operators symmetric and traceless on $A,B$, for example). This
suggests that the interaction terms in the string worldsheet
Hamiltonian should not involve zero-mode oscillators at all. We
will eventually see that this is the case, at least to the order
we are able to study.

We have given a rather detailed treatment of the calculation of
the anomalous dimensions of two specific operator multiplets. To
fully address the issues that will arise in string theory, we need
expressions like (\ref{finlopdim}) for {\it all} operator
multiplets (not just spacetime scalars) that contain components
with $\Delta_0 = 2$. It is possible to carry out some version of
the above lattice Laplacian argument for all the relevant operator
classes, but we can use supersymmetry to circumvent this tedious
task. The extended superconformal symmetry of the gauge theory
means that conformal primary operators are organized into
multiplets obtained from a lowest-dimension primary ${\cal O}_D$
of dimension $D$ by anticommutation with the supercharges
$Q_i^\alpha$ ($i$ is an $SL(2,C)$ Lorentz spinor index and
$\alpha$ is an $SU(4)$ index). We need only concern ourselves here
with the case in which ${\cal O}_D$ is a spacetime scalar (of
dimension $D$ and $R$-charge $R$). There are sixteen supercharges
and we can choose eight of them to be raising operators; there are
$2^8=256$ operators we can reach by `raising' the lowest one.
Since the raising operators increase the dimension and $R$-charge
by 1/2 each time they act, the operators at level L, obtained by
acting with L supercharges, all have the same dimension and
$R$-charge. The corresponding decomposition of the 256-dimensional
multiplet is shown in table~\ref{table1}.

\begin{table}[ht!]
\begin{eqnarray}
\begin{array}{|l|l|l|l|l|l|l|l|l|l|}\hline
{\rm Level}& 0& 1& 2& 3& 4& 5& 6& 7& 8 \\ \hline {\rm
Multiplicity}& 1& 8& 28& 56& 70& 56& 28& 8& 1 \\ \hline {\rm
Dimension}& D& D+{1}/{2} & D+{1}& D+{3}/{2} & D+{2} & D+{5}/{2} &
D+{3} & D+{7}/{2} & D+4 \\ \hline R-{\rm charge} & R& R+{1}/{2} &
R+{1}& R+{3}/{2} & R+{2} & R+{5}/{2} & R+{3} & R+{7}/{2} & R+4 \\
\hline
\end{array} \nonumber
\end{eqnarray}
\caption{$R$-charge content of a supermultiplet} \label{table1}
\end{table}
\medskip
\noindent The states at each level can be classified under the
Lorentz group and the $SO(4)\sim SU(2)\times SU(2)$ subgroup of
the $R$-symmetry group, which is unbroken after we have fixed the
$SO(2)$ $R$-charge. For instance, the 28 states at level 2
decompose under $SO(4)_{Lor}\times SO(4)_R$ as
$(6,1)+(1,6)+(4,4)$. For the present, the most important point is
that, given the dimension of one operator at one level, we can
infer the dimensions of all other operators in the supermultiplet.

We can use this logic to get a complete accounting of the
dimensions of the $\Delta_0=2$ BMN operators. Here we summarize
work by Beisert \cite{BeisertSUSY}, recasting his results to fit
our needs (and adding some further useful information that emerges
from our own $SU(4)$ analysis). The supermultiplet of interest is
based on the set of scalars
$\Sigma_A\tr\left(\phi^AZ^p\phi^AZ^{R-p}\right)$, the operator
class we have denoted by $\overline T_{R+2}^{(0)}$. According to
(\ref{firstopdim}), the spectrum of $\Delta=D-R$ eigenvalues
associated with this operator basis is
\begin{eqnarray}
\label{multopdim}
\Delta(\overline T_{R+2}^{(0)}) =
    2 + \frac{g^2_{YM} N_c}{\pi^2} \sin^2\left(\frac{n\pi}{R+3}\right)
    \rightarrow 2 +\frac{g^2_{YM} N_c}{R^2}~n^2~\left(1-\frac{6}{R}
+O(R^{-2})\right)~.
\end{eqnarray}
The other spacetime scalar operators $\overline T_{R+2}^{(\pm)}$
displayed in (\ref{firstopdim}) have dimension formulae which
appear to differ from this. However, when they are put into the
context of a supermultiplet and the dimension formulae are
expressed in terms of the $R$-charge of the lowest-dimension
member of the supermultiplet, it turns out that (\ref{multopdim})
governs {\it all} the operators at {\it all} levels in the
supermultiplet. We summarize the situation for the spacetime
scalar members of the multiplet in table~\ref{tableone}.
\begin{table}[ht!]
\begin{eqnarray}
\begin{array}{|l|l|l|l|l|l|}\hline
L& R& SU(4) ~ {\rm Irreps} & {\rm Operator} & \Delta-2 & {\rm Multiplicity}
\\ \hline
0 & R_0 & (0,R_0,0) & \Sigma_A\tr\left(\phi^AZ^p\phi^AZ^{R_0-p}\right) &
\frac{g^2_{YM} N_c}{\pi^2} \sin^2(\frac{n\pi}{(R_0)+3}) &
    n=1,.,\frac{R_0+1}{2} \\ \hline
2 & R_0+1 & (0,R_0,2)+c.c. & \tr\left(\phi^{[i}Z^p\phi^{j]}Z^{R_0+1-p}\right) &
\frac{g^2_{YM} N_c}{\pi^2} \sin^2(\frac{n\pi}{(R_0+1)+2}) &
    n=1,.,\frac{R_0+1}{2} \\ \hline
4 & R_0+2 & (2,R_0,2) & \tr\left(\phi^{(i}Z^p\phi^{j)}Z^{R_0+2-p}\right) &
\frac{g^2_{YM} N_c}{\pi^2} \sin^2(\frac{n\pi}{(R_0+2)+1}) &
    n=1,.,\frac{R_0+1}{2} \\ \hline
4&R_0+2&(0,R_0+2,0)\times 2&
\tr\left(\chi^{[\alpha}Z^p\chi^{\beta]}Z^{R_0+1-p}\right) &
\frac{g^2_{YM} N_c}{\pi^2} \sin^2(\frac{n\pi}{(R_0+2)+1}) &
    n=1,.,\frac{R_0+1}{2} \\ \hline
6&R_0+3&(0,R_0+2,2)+c.c.&\tr\left(\chi^{(\alpha}
Z^p\chi^{\beta)}Z^{R_0+2-p}\right) &
\frac{g^2_{YM} N_c}{\pi^2} \sin^2(\frac{n\pi}{(R_0+3)+0}) &
    n=1,.,\frac{R_0+1}{2} \\ \hline
8 & R_0+4 & (0,R_0,0) & \tr\left(\nabla_\mu Z Z^p\nabla^\mu Z Z^{R_0+2-p}
\right) &
\frac{g^2_{YM} N_c}{\pi^2} \sin^2(\frac{n\pi}{(R_0+4)-1}) &
    n=1,.,\frac{R_0+1}{2} \\ \hline
\end{array}
\nn\\ \nonumber
\end{eqnarray}
\caption{Dimensions and multiplicities of spacetime scalar
operators} \label{tableone}
\end{table}
\noindent The last column displays the allowed range of the
eigenvalue index $n$ at each level (for $R_0$ odd only, just to save
space) computed from our results for $SU(4)$ irrep multiplicities.
It is non-trivial that the result is the same at each level; were
it not so, the levels could not be assembled into a single
supermultiplet. The universal dimension formula is written at each
level in such a way as to emphasize the dependence on the
$R$-charge of the particular level. This shows how the different
results (\ref{firstopdim}) and (\ref{multopdim}) are reconciled in
the supermultiplet.

The supermultiplet contains operators that are not spacetime
scalars (i.e., that transform non-trivially under the $SU(2,2)$
conformal group) and group theory determines at what levels in
the supermultiplet they must lie. A representative sampling of
data on such operators (extracted from Beisert's paper) is
collected in table~\ref{tabletwo}. We have worked out neither the
$SU(4)$ representations to which these lowest-$\Delta$ operators
belong nor their precise multiplicities. The ellipses indicate
that the operators in question contain further monomials involving
fermion fields (so that they are not uniquely specified by their
bosonic content). This information will be useful in consistency
checks to be carried out below.

\begin{table}[ht!]
\begin{eqnarray}
\begin{array}{|l|l|l|l|l|}\hline
L & R& {\rm Operator} & \Delta-2 &  \Delta-2\to \\ \hline 2 & R_0+1 &
\tr\left(\phi^iZ^p\nabla_\mu Z Z^{R_0-p}\right)+\ldots &
\frac{g^2_{YM} N_c}{\pi^2} \sin^2(\frac{n\pi}{(R_0+1)+2}) &
\frac{g^2_{YM} N_c}{R_0^2}n^2(1-\frac{4}{R_0}) \\ \hline 4 & R_0+2  &
\tr\left(\phi^iZ^p\nabla_\mu Z Z^{R_0+1-p}\right) & \frac{g^2_{YM}
N_c}{\pi^2} \sin^2(\frac{n\pi}{(R_0+2)+1}) & \frac{g^2_{YM}
N_c}{R_0^2}n^2(1-\frac{2}{R_0}) \\ \hline 4 & R_0+2  &
\tr\left(\nabla_{(\mu}ZZ^p\nabla_{\nu)}Z Z^{R_0-p}\right) &
\frac{g^2_{YM} N_c}{\pi^2} \sin^2(\frac{n\pi}{(R_0+2)+1}) &
\frac{g^2_{YM} N_c}{R_0^2}n^2(1-\frac{2}{R_0}) \\ \hline 6 & R_0+3  &
\tr\left(\phi^iZ^p\nabla_\mu Z Z^{R_0+2-p}\right)+\ldots &
\frac{g^2_{YM} N_c}{\pi^2} \sin^2(\frac{n\pi}{R_0+3}) &
\frac{g^2_{YM} N_c}{R_0^2}n^2(1-\frac{0}{R_0}) \\ \hline 6 & R_0+3  &
\tr\left(\nabla_{[\mu}ZZ^p\nabla_{\nu]}Z Z^{R_0+1-p}\right) &
\frac{g^2_{YM} N_c}{\pi^2} \sin^2(\frac{n\pi}{R_0+3}) &
\frac{g^2_{YM} N_c}{R_0^2}n^2(1-\frac{0}{R_0}) \\ \hline
\end{array}\nonumber
\end{eqnarray}
\caption{Anomalous dimensions of some operators that are not
scalars} \label{tabletwo}
\end{table}

As far as dimensions are concerned, all of the above can be
summarized by saying that the dimensions of the operators of
$R$-charge $R$ at level $L$ in the supermultiplet are given by the
general formula (valid for large $R$ and fixed $n$):
\begin{eqnarray}
\label{deltalevel} \Delta^{R,L}_n = 2 + \frac{g^2_{YM} N_c}{\pi^2}
\sin^2\left(\frac{n\pi}{R+3-L/2}\right) \to 2+\frac{g^2_{YM}
N_c}{R^2}~n^2\left(1-\frac{6-L}{R}+O(R^{-2})\right)~.
\end{eqnarray}
This amounts to a gauge theory prediction for the way in which
worldsheet interactions lift the degeneracy of the two-impurity
string multiplet. The 256 states of the form $A^\dagger_n
B^\dagger_{-n} \vert R\rangle$, for a given mode number $n$,
(where $A^\dagger$ and $B^\dagger$ each can be any of the 8+8
bosonic and fermionic oscillators) should break up as shown in
table~\ref{smultiplicity}. It should be emphasized that, for fixed
$R$, the operators associated with different levels are actually
coming from {\it different} supermultiplets; this is why they have
different dimensions! As mentioned before, we can also precisely
identify transformation properties under the Lorentz group and
under the rest of the $R$-symmetry group of the degenerate states
at each level. This again leads to useful consistency checks, and
we will elaborate on this when we analyze the eigenstates of the
string worldsheet Hamiltonian.
\begin{table}[ht!]
\begin{eqnarray}
\begin{array}{|l|l|l|l|l|l|l|l|l|l|}\hline
{\rm Level} & 0& 1& 2& 3& 4& 5& 6& 7& 8 \\ \hline
{\rm Multiplicity} & 1& 8& 28& 56& 70& 56& 28& 8& 1 \\ \hline
\delta E\times (R^2/g_{YM}^2N_c n^2) & -{6}/{R} & -{5}/{R} &
-{4}/{R} & -{3}/{R} & -{2}/{R} & -{1}/{R} &
0  & {1}/{R} & {2}/{R} \\ \hline
\end{array} \nonumber
\end{eqnarray}
\caption{Predicted energy shifts of two-impurity string states}
\label{smultiplicity}
\end{table}

\section{Worldsheet Action, Curvature Expansion, Light-Cone Reduction}

We now turn to the construction of the classical GS superstring
action in the $AdS_5\times S^5$ target space. We would like to
construct a worldsheet action that has the full $SO(4,2)\times
SO(6)$ symmetry of this space (we speak only about the bosonic
symmetries, but similar considerations apply to their fermionic
partners as well). However, it is not possible to make the full
symmetry manifest: the fact that one has to expand about a
classical trajectory of the string `spontaneously breaks' the
symmetry down to $SO(4,1)\times SO(5)$ (the analog of the
Poincar\'e group for this background). The solution to this
problem lies in the fact that this target space can be realized as
the coset space $SO(4,2)\times SO(6)/SO(4,1)\times SO(5)$: there
is a general strategy for writing a nonlinear sigma model action
on a target space $G/H$ such that only symmetry under the
stabilizer group $H$ is manifest (i.e. linearly realized) while
the remaining generators of the full symmetry group $G$ are
realized nonlinearly. This coset construction can be generalized
to handle supersymmetries as well, provided that the target
superspace can be realized as a supercoset manifold. Fortunately,
this is true for the superstring on $AdS_5\times S^5$, as was
shown by Metsaev and Tseytlin \cite{MetTseyt} who constructed an
action possessing the full $PSU(2,2|4)$ supersymmetry. Their
action consists of a kinetic term and a Wess--Zumino term built as
follows out of Cartan (super)one-forms on the supercoset manifold:
\begin{eqnarray} \label{GSactone}
2\pi\ap {\cal S}_{GS}& = & \int d^2\sigma\left( {\cal L}_{\rm Kin} +
        {\cal L}_{\rm WZ} \right) \nn\\
& = & \int d^2\sigma\left(-\frac{1}{2} h^{ab} L_a^\mu L_b^\nu\eta_{\mu\nu}\
 -2i\epsilon^{ab} \int_0^1 dt\, L_{at}^\mu s^{IJ}
    \bar\theta^I \Gamma^\nu L_{bt}^J\eta_{\mu\nu}\right)\ .
\nn\\
\end{eqnarray}
The $\Gamma^\mu$ are $SO(9,1)$ gamma matrices,
$\eta_{\mu\nu}$ is the $SO(9,1)$ Minkowski metric and 
$s^{IJ}={\rm diag}(1,-1)$.
The world-sheet fermi fields $\theta^I$ ($I,J=1,2$)
of the type IIB theory are two
$SO(9,1)$ Majorana--Weyl spinors of the same chirality satisfying
$\Gamma_{11}\theta^I=\theta^I$. 
The gauge-fixing condition
$\Gamma_0\Gamma_9\theta^I=\theta^I$ allows us to set half the
components of $\theta^I$ to zero. This condition will be kept
exact throughout the gauge-fixing and curvature expansion
procedure (as will the bosonic gauge condition $x^+=\tau$). It
will also be convenient to define a complex spinor $\psi =
\sqrt{2}(\theta^1 + i\theta^2)$.  Upon restricting to the ${\bf
8}_s$ representation, the condition $\gamma^9\psi = +\psi$ selects
the upper eight components of $\psi$, since $\gamma^9 = {\rm
diag}(1,-1)_{16\times 16}$.  The fermion $\psi_\alpha$ can
therefore be thought of as an eight-component complex spinor
constructed from the 16 components of $\theta^I$ that survive the
above gauge fixing.

The Cartan one-forms satisfy constraint equations, known
as the Maurer--Cartan equations, which can be thought of as
generalized Bianchi identities.  
The approach in \cite{MetTseyt} was to solve these equations
order-by-order in powers of the coordinate fields $(x,\theta)$,
and the first few terms (to quartic order in fields) 
of (\ref{GSactone}) were written explicitly therein.
Following \cite{MetTseyt},
it was shown by Kallosh, Rahmfeld and Rajaraman
that these equations can be solved exactly for
the $AdS_5\times S^5$ geometry with the following results
\cite{Kallosh:1998zx}:
\begin{eqnarray}
\label{sol}
L_{bt}^J  =  \frac{\sinh t{\cal M}}{{\cal M}} {\cal
D}_b \theta^J
    & \qquad &
L_{at}^\mu  =  e^\mu_{\phantom{\mu}\rho}\partial_a x^\rho
    - 4i\bar\theta^I \Gamma^\mu
    \left( \frac{\sinh^2 (t{\cal M}/2)}{{\cal M}^2}
    \right){\cal D}_a \theta^I\ , \nn \\
    L_{b}^J =L_{bt}^J~\vert_{t=1} & \qquad &
        L_{a}^\mu =L_{at}^\mu~\vert_{t=1}\ ,
\end{eqnarray}
where
\begin{eqnarray}
\label{}
({\cal D}_a \theta)^I
    & = & \left( \partial_a \theta + {1\over 4}
  \left(\omega^{\mu\,\nu}_{\phantom{\mu\,\nu}\,\rho}\,\partial_a x^\rho \right)
    \Gamma^{\mu\nu} \theta\right)^I
  -{i\over 2}\epsilon^{IJ} e^\mu_{\phantom{\mu}\,\rho}\,\partial_a x^\rho
    \Gamma_* \Gamma^\mu \theta^J \ , \nn \\
({\cal M}^2)^{IL} & = &
  -\epsilon^{IJ}(\Gamma_* \Gamma^\mu \theta^J \bar\theta^L \Gamma^\mu)
 +\frac{1}{2}\epsilon^{KL}(-\Gamma^{jk}
 \theta^I \bar\theta^K \Gamma^{jk}\Gamma_*
      + \Gamma^{j'k'}\theta^I \bar\theta^K \Gamma^{j'k'} {\Gamma'}_*)\ .
\end{eqnarray}
In the above, $e^\mu_{\phantom{\mu}\,\rho}$
($\omega^{\mu\,\nu}_{\phantom{\mu\,\nu}\,\rho}$) is the vielbein
(spin connection) in the $AdS_5\times S^5$ geometry, and $\Gamma_*
\equiv i\Gamma_{01234}$, ${\Gamma'}_* \equiv i\Gamma_{56789}$.
Factors of the Minkowski metric, needed to contract Lorentz indices,
have been suppressed.

The action obtained by substituting these formulas into (\ref{GSactone}) can
now be systematically expanded in powers of the inverse curvature
scale $\widehat R$ by inserting the expansions of the vielbeins
and spin connections that follow from the metric expansion
(\ref{expndmet}). The expansion of $L_a^\mu L_b^\mu$, from which
one constructs the kinetic term of the worldsheet action, is
\begin{eqnarray}
\label{LaLb} L_0^\mu L_0^\mu & \approx & \left\{ 2{}\dot x^- -
(x^A)^2 + (\dot x^A)^2
    -2i{}\bar\theta^I\Gamma^-
    (\partial_0\theta^I-{}\epsilon^{IJ}\Pi\theta^J)\right\}
\nonumber \\ & &     + \frac{1}{\Rhat^2}\biggl\{ (\dot x^-)^2 -
2{}y^2\dot x^- + \frac{1}{2}(\dot z^2 z^2 - \dot y^2 y^2)
        + \frac{1}{2}(y^4-z^4) + (\theta ~{\rm terms}) \biggr\} \nn \\
L_1^\mu L_1^\mu & \approx &
        ({x'}^A)^2 + \frac{1}{\Rhat^2}\left\{
        \frac{1}{2}({z'}^2 z^2 - {y'}^2 y^2)
        + ({x'}^-)^2 + (\theta ~ {\rm terms}) \right\} \nn \\
L_0^\mu L_1^\mu & \approx & \left\{ {x'}^- + \dot x^A{x'}^A
    -i{} \bar\theta^I\Gamma^-\partial_1\theta^I\right\} \nn \\
& &     + \frac{1}{\Rhat^2}\biggl\{ {x'}^- \dot x^- - {}y^2{x'}^-
+
  \frac{1}{2}(z^2 \dot z_k z'_k -y^2 \dot y_{k'}{y'}_{k'})
    + (\theta ~ {\rm terms}) \biggr\}\ ,
\end{eqnarray}
The bosonic coordinate $x^A = (y_{k'},z_k)$ has eight components.
(In the previous section the indices $A,B$ took four values, but
in this section they take eight values.) As usual, a dot is the
same thing as $\partial_0$ and a prime is the same thing as
$\partial_1$. The expression for the full worldsheet action
expanded in this fashion is not very illuminating, and we will not
present it here.

Our ultimate goal is to construct a Hamiltonian for the physical
transverse coordinates $x^A, ~\psi^\alpha$ and their associated
canonical momenta. The first step is to impose the bosonic gauge
condition $x^+ = \tau$ along with the $\kappa$-symmetry
gauge-fixing condition $\Gamma^+ \theta^I = 0$. At leading order
in $1/\Rhat$, the gauge $x^+ =\tau$ is consistent with a flat
worldsheet metric $h^{ab}=(-1,1)$. However, to maintain the gauge
choice $x^+ =\tau$ beyond leading order, it turns out that we must
allow $h^{ab}$ to acquire curvature corrections.\footnote{This is
to be contrasted with ref.~\cite{Parnachev:2002kk}, which imposed
a flat world sheet metric and introduced curvature corrections to
the gauge choice.} We therefore need to eliminate both $x^-$ and
the worldsheet metric in favor of physical variables. Taken
together, the equations of motion for $x^-$ and the conformal
gauge constraints (vanishing of the worldsheet energy-momentum
tensor) provide exactly the information needed to do this.

Consider first the conformal constraints obtained by varying the
action with respect to the worldsheet metric:
\begin{eqnarray}
\label{enmomten}
T_{ab} = L^\mu_a L^\mu_b -\frac{1}{2} h_{ab}
h^{cd} L^\mu_c L^\mu_d = 0~.
\end{eqnarray}
Because $T_{ab}$ is symmetric
and traceless, there are only two independent constraints
associated with conformal invariance on the worldsheet. Using
(\ref{LaLb}) and (\ref{expndmet}), and keeping only terms of
leading order in $1/\Rhat$, they read
\begin{eqnarray} \label{Tab}
T_{00} & \sim & \frac{1}{2}\left( 2{}\dot x^- - (x^A)^2 + (\dot
x^A)^2
            + ({x'}^A)^2 -2i{}\bar\theta^I\Gamma^-
        (\partial_0\theta^I-{}\epsilon^{IJ}\Pi\theta^J)
    \right)  = 0 \nn \\
T_{01} & \sim & {x'}^- + \dot x^A {x'}^A
    -i{} \bar\theta^I\Gamma^-\partial_1\theta^I = 0~.
\end{eqnarray}
These constraints can be recast as equations to determine $x^-$ to
leading order in $1/\Rhat$:
\begin{eqnarray}
\label{xminus} (\dot x^-)_0 & = & \frac{1}{2}(x^A)^2 -
\frac{1}{2{}}\left[(\dot x^A)^2 +
        ({x'}^A)^2\right] + i\bar\theta^I\Gamma^-
        (\partial_0\theta^I-{}\epsilon^{IJ}\Pi\theta^J) \nn \\
({x^\prime}^-)_0 & = & - \dot x^A {x'}^A
    +i \bar\theta^I\Gamma^-\partial_1\theta^I ~.
\end{eqnarray}
We will eventually show how to evaluate the $O(1/\Rhat^2)$ corrections
to $x^-$.

The building blocks of the $x^-$ equation of motion, to
$O(1/\Rhat^2)$, are as follows:
\begin{eqnarray}
\label{xmineom} \frac{\delta {\cal L}}{\delta \dot x^-} & = &
\frac{1}{2}h^{00} \left\{ {2{}} + \frac{1}{\Rhat^2}\left[2\dot x^-
        -2{}y^2 - i\bar\theta^I\Gamma^- \partial_0 \theta^I
      +2i{}\bar\theta^I\Gamma^-
      \epsilon^{IJ}\Pi\theta^J\right]\right\}
+\frac{i}{2\Rhat^2}s^{IJ}\bar\theta^I\Gamma^-\partial_1\theta^J
\nn \\ & = &  \frac{1}{2}h^{00}\left\{ {2{}}
        + \frac{1}{\Rhat^2}\left[{}(z^2-y^2)
    - \left[(\dot x^A)^2 + ({x'}^A)^2\right]
        + i\bar\theta^I\Gamma^-\partial_0\theta^I \right] \right\}
        + \frac{i}{2\Rhat^2}s^{IJ}
\bar\theta^I\Gamma^-\partial_1\theta^J \nn\\ \frac{\delta {\cal
L}}{\delta {x'}^-} & = & {h^{01}{}}
    + \frac{h^{11}}{\Rhat^2}\left( - \dot x^A {x'}^A
        + \frac{i}{2}\bar\theta^I\Gamma^- \partial_1\theta^I \right)
        -\frac{i}{2\Rhat^2}s^{IJ}\bar\theta^I\Gamma^-\partial_0\theta^J~.
\end{eqnarray}
In the first equation, $x^-$ was eliminated by using the
$T_{00}$ constraint evaluated to leading order. It is obvious from
(\ref{xmineom}) that the choice of a flat Minkowski worldsheet
metric ($h^{00}=-h^{11}=1, h^{01}=0$) is inconsistent with the
$x^-$ equations of motion. To allow for corrections to the metric,
we therefore write
\begin{equation}
h^{00}  =  -1 + \frac{\tilde h^{00}}{\Rhat^2} + \dots \qquad
h^{11}  =  1 + \frac{\tilde h^{11}}{\Rhat^2} + \dots \qquad
h^{01}  =  \frac{\tilde h^{01}}{\Rhat^2} + \dots~.
\end{equation}
By choosing the metric corrections
\begin{eqnarray}
\label{h00}
\tilde h^{00} & = & \frac{1}{2}(z^2-y^2)
        - \frac{1}{2}\left[(\dot x^A)^2 + ({x'}^A)^2\right]
        + \frac{i}{2{}}\bar\theta^I\Gamma^-\partial_0\theta^I
        -\frac{i}{2{}}s^{IJ}\bar\theta^I\Gamma^-\partial_1\theta^J
\end{eqnarray}
\begin{eqnarray}
\label{h01} \tilde h^{01} & = & \dot x^A {x'}^A
        - \frac{i}{2{}}\bar\theta^I \Gamma^- \partial_1\theta^I
        +\frac{i}{2{}}s^{IJ}\bar\theta^I\Gamma^-\partial_0\theta^J\ ,
\end{eqnarray}
the $x^-$ equation of motion is vastly simplified:
\begin{eqnarray}
\frac{\delta {\cal L}}{\delta \dot x^-} = 1 + O(1/\Rhat^4) \qquad
\frac{\delta {\cal L}}{\delta {x'}^-} =  O(1/\Rhat^4) .
\end{eqnarray}
This choice of worldsheet metric is what is needed to enforce the
lightcone gauge condition $x^+ =\tau$ to $O(1/\Rhat^2)$. With the
corrected metric in hand, we can revisit the conformal gauge
constraints to determine $x^-$ to $O(1/\Rhat^2)$.

Upon evaluating $x^-$ to the order of interest, we are equipped to
express the Hamiltonian density for the generator of translations
of lightcone time $x^+$, $\delta{\cal L}/\delta\dot x^+= p_{+}$.
The variation is done before any gauge fixing, holding the
remaining coordinates and the worldsheet metric fixed. The
replacement of $x^\pm$ and $h^{ab}$ by fixing conformal gauge is
understood to be completed after the variation. The result to
$O(1/\Rhat^2)$ is
\begin{eqnarray}
{\cal H}_{lc}& = & \frac{\delta {\cal
L}_{\rm GS}}{\delta \dot x^+} =
        ~ {\cal H}_{pp} + {\cal H}_{int} \nn \\
{\cal H}_{pp}& = &\frac{1}{2}(x^A)^2 + \frac{1}{2{}}\left[(\dot
x^A)^2
    + ({x'}^A)^2\right] -i{}\bar\theta^I\Gamma^-\epsilon^{IJ}\Pi\theta^J
        +is^{IJ}\bar\theta^I\Gamma^-\partial_1\theta^J \nn
\end{eqnarray}
\begin{eqnarray}
\label{GSlchamdens}
\widehat R^2 {\cal H}_{int} & = &
        \frac{1}{4{}}\left[y^2(\dot z^2 - {z'}^2 - 2{y'}^2)+
        z^2(-\dot y^2 + {y'}^2+2{z'}^2) \right]
\nonumber \\ & &     +\frac{1}{8}\left[3(\dot
x^A)^2-({x'}^A)^2\right]
        \left[(\dot x^A)^2 + ({x'}^A)^2\right]
        +\frac{1}{8}\left[(x^A)^2\right]^2
        -\frac{1}{2}(\dot x^A{x'}^A)^2
\nonumber \\ & &    -\frac{i}{4{}}\sum_{a=0}^1\bar\theta^I
(\partial_a x^A \Gamma^A )\epsilon^{IJ}\Gamma^-
        \Pi (\partial_a x^B \Gamma^B)\theta^J
        -\frac{i}{2}{}(x^A)^2\bar\theta^I\Gamma^-\epsilon^{IJ}\Pi\theta^J
\nonumber \\ & &     -\frac{i}{2}(\dot
x^A)^2\bar\theta^I\Gamma^-\partial_0\theta^I
        -\frac{i{}}{12}\bar\theta^I\Gamma^-({\cal M}^2)^{IJ}
        \epsilon^{JL}\Pi\theta^L
        -\frac{1}{2}(\bar\theta^I\Gamma^-\epsilon^{IJ}\Pi\theta^J)^2
\nonumber \\ & &     -\frac{i}{2}(\dot x^A
{x'}^A)s^{IJ}\bar\theta^I \Gamma^-\partial_0\theta^J
        -\frac{i}{4}(y^2-z^2)s^{IJ}\bar\theta^I\Gamma^-\partial_1\theta^J
\nonumber \\
& &     +\frac{i}{4}{x'}^A s^{IJ}\bar\theta^I
\Gamma^A(y_{j'}\Gamma^{-j'}-z_j\Gamma^{-j})\theta^J
        +\frac{i}{4}s^{IJ}\bar\theta^I\Gamma^-(z'_j z_k
        \Gamma^{jk} - y'_{j'}y_{k'}\Gamma^{j'k'})\theta^J
\nonumber \\ & &     +\frac{i}{4}\left[(\dot
x^A)^2-({x'}^A)^2\right]
s^{IJ}\bar\theta^I\Gamma^-\partial_1\theta^J
        +\frac{i}{12}s^{IJ}\bar\theta^I
        \Gamma^-({\cal M}^2)^{JL}\partial_1\theta^L
\nonumber \\
& &     +\frac{1}{2}(s^{IJ}\bar\theta^I\Gamma^-\partial_1\theta^J)
(\bar\theta^K\Gamma^-\epsilon^{KL}
        \Pi\theta^L) + \frac{i}{4}(x^A)^2 s^{IJ}\bar\theta^I
        \Gamma^-\partial_1\theta^J\ .
\end{eqnarray}

To quantize this system, the Hamiltonian must be expressed in
terms of canonical coordinates and momenta. The recipe for
computing the bosonic momenta $p^A$ is, once again, to vary ${\cal
L}$ with respect to $\dot x^A$, holding the other coordinates and
the worldsheet metric fixed, and replacing $x^\pm$ and $h^{ab}$
according to the appropriate constraints only after the variation
is done. For example, the result for the momenta in the $SO(4)$
descending from $AdS_5$, correct to $O(1/\Rhat^2)$, is
\begin{eqnarray}
\label{bosomom}
p_k & = & \dot z_k +
\frac{1}{\Rhat^2}\biggl\{\frac{1}{2} y^2 p_k
    + \frac{1}{2}\left[ (p_A)^2 + ({x'}^A)^2 \right]p_k
        - (p_A {x'}^A ){z'}_k
       - \frac{i}{2{}}p_k \bar\theta^I \Gamma^-\partial_0\theta^I \nn\\
& &     + \frac{i}{2{}}p_k
s^{IJ}\bar\theta^I\Gamma^-\partial_1\theta^J
        - \frac{i{}}{4}\bar\theta^I\Gamma^-
        z_j \Gamma_k^{\phantom{k}j}\theta^I
        + \frac{i{}}{4}\bar\theta^I \Gamma^k 
	y_{j'}\Gamma^{-j'}  \theta^I
	+ \frac{i}{2{}}{z'}_k\bar\theta^I\Gamma^-\partial_1 \theta^I
\nn\\
& &     + \frac{i}{4}p_A \epsilon^{IJ}\bar\theta^I\Gamma^- \left(
      \Gamma_k \Pi \Gamma^A  + \Gamma^A \Pi \Gamma_k \right ) \theta^J
        - \frac{i}{2{}}{z'}_k s^{IJ}
        \bar\theta^I\Gamma^-\partial_0\theta^J \nn\\
& &     + \frac{i}{4}{x'}^A s^{IJ}\epsilon^{JK}\bar\theta^I\Gamma^-
    \left( \Gamma_k\Pi\Gamma^A - \Gamma^A \Pi
    \Gamma_k \right) \theta^K \biggr\}\ .
\end{eqnarray}
To calculate the Hamiltonian to $O(1/\widehat R^{2})$, we use this
relation to eliminate $\dot z^k$ from ${\cal H}_{pp}$ (and use
$p_k=\dot z^k$ in ${\cal H}_{int}$).

Performing the analogous operation in the fermionic regime is more
complicated. At this point it is convenient to change notation by
replacing fermionic coordinates $\theta^I$ with the single complex
spinor $\psi=\sqrt{2}(\theta^1+\imath\theta^2)$; $\psi$ and
$\psi^\dagger$ appear as independent coordinates and there are two
fermionic canonical momenta, $\rho_\psi=\delta{\cal
L}/\delta\dot\psi$ and $\rho_{\psi^\dagger}= \delta{\cal
L}/\delta\dot\psi^\dagger$. In all the standard field theory
examples, one can manipulate ${\cal L}$ (using integration by
parts in time) so that the action is independent of
$\dot\psi^\dagger$. The momentum equations, known as the {\it
primary constraints}, then read $\rho_\psi-\psi^\dagger=0$ and
$\rho_{\psi^\dagger}=0$. They are, in effect, constraints that
eliminate $\psi^\dagger$ and $\rho_{\psi^\dagger}$ as dynamical
variables in the system, leaving the standard canonical Poisson
brackets for $\psi, \rho_\psi$ unchanged. Things do not work quite
so simply in the present problem.

The terms in ${\cal L}_{GS}$ that depend on $\dot\psi$ and
$\dot\psi^\dagger$ (and are therefore relevant for the fermionic
momentum constraints) are
\begin{eqnarray}
\label{Lfermimom} {\cal L} & \sim & -i{} \left( \psi^\dagger
\dot\psi \right)
        - \frac{i}{\Rhat^2}\biggl\{
        \frac{1}{4}\left[\dot x^- + \frac{1}{2}(z^2 - y^2)\right]
                \left(\psi\dot \psi^\dagger + \psi^\dagger\dot\psi\right)
        -\frac{{}\tilde h^{00}}{2}\left(\psi\dot\psi^\dagger
    + \psi^\dagger\dot\psi\right) \nn\\
& &     + \frac{1}{96}\left(\psi\gamma^{jk}\psi^\dagger\right)
        \left(\psi\gamma^{jk}\Pi\dot\psi^\dagger
    - \psi^\dagger\gamma^{jk}\Pi\dot\psi\right)
       -\frac{{x'}^-}{4}\left(\psi\dot\psi 
	+ \psi^\dagger\dot\psi^\dagger\right)
        - (j,k \rightleftharpoons j',k') \biggr\} .
\end{eqnarray}
Some of the curvature correction terms in (\ref{Lfermimom}) contain
$\dot\psi^\dagger$ in such a way that it cannot be eliminated from the
action. The fermionic constraints, or primary constraints, therefore take
on a more complicated form than usual:
\begin{eqnarray}
\label{rhoeqns} \rho_\psi & = &   i{} \psi^\dagger +
\frac{1}{\Rhat^2}\biggl\{
 \frac{1}{4}y^2\rho + \frac{1}{8}\left[ (p_A^2) + ({x'}^A)^2\right] \rho
        + \frac{i}{4{}}(p_A {x'}^A )\psi
        + \frac{i}{4{}}\left( \rho\Pi\psi \right) \rho
\nn\\ & &     - \frac{i}{8{}}\left( \psi\rho' + \rho\psi'
\right)\psi
        + \frac{i}{8{}}\left(\psi\psi'
         - \rho\rho'\right)\rho
\nn\\ & &     + \frac{i}{48{}}\left[
\left(\psi\gamma^{jk}\rho\right) \left(\rho\gamma^{jk}\Pi\right)
        - (j,k, \rightleftharpoons j',k') \right]
        \biggr\}
\\
\rho_{\psi^\dagger} & = & \frac{1}{\Rhat^2}\biggl\{ \frac{i}{4}{}
y^2\psi + \frac{i}{8{}} \left[ (p_A^2) + ({x'}^A)^2 \right]\psi
        + \frac{1}{4}\left( p_A {x'}^A \right)\rho
        -\frac{1}{4}\left( \rho\Pi\psi \right)\psi
\nn\\ & &     -\frac{1}{8}\left(\psi\rho' + \rho\psi' \right)\rho
        - \frac{1}{8}\left(\psi\psi'
        - \rho\rho'\right)\psi
        \biggr\} .
\end{eqnarray}
Again, this is the result of varying the Lagrangian first and then substituting
in solutions for the gauge-fixed coordinates and worldsheet metric.  For
clarity, we express the primary constraints as follows:
\begin{eqnarray}
\label{primary} \chi_\psi & \equiv & \rho_\psi - i{}\psi^\dagger -
\left(
    1/\Rhat^2\ {\rm corrections} \right) = 0 \nn\\
\chi_{\psi^\dagger} & \equiv & \rho_{\psi^\dagger} - \left(
    1/\Rhat^2\ {\rm corrections} \right) = 0\ .
\end{eqnarray}

We are now ready to proceed with rewriting the Hamiltonian as a
function of canonical coordinates and momenta. The elimination of
the bosonic velocities via the equations that define the bosonic
canonical momenta (\ref{bosomom}) is straightforward, at least to
first non-leading order in $1/\Rhat^2$. Although it requires a
messy calculation to show it, this step also eliminates all terms
involving $\dot\psi$ or $\dot\psi^\dagger$. As a result, all terms
involving fermi fields are built out of $\psi$, $\psi^\dagger$ and
their $\sigma$ derivatives. It is again straightforward, at least
to first subleading order in $1/\Rhat^2$, to use the $\rho_\psi$
constraint to eliminate $\psi^\dagger$ in favor of $\rho_\psi$.
The result is exactly what we want: a Hamiltonian expressed as a
function of $x^A,p^A$ and $\psi^\alpha,\rho^\alpha_\psi$. We are
not quite done, however. In general, a set of primary constraints
$\chi=0$ (\ref{primary}) can be categorized as either first or
second-class constraints.  The so-called second-class constraints
arise when canonical momenta (defined by the primary constraints)
do not have vanishing Poisson brackets with the primary
constraints themselves: $\left\{ \rho_\psi,\chi_\psi \right\} \neq
0$, $\left\{ \rho_{\psi^\dagger},\chi_{\psi^\dagger} \right\} \neq
0$. (First-class constraints are characterized by the more typical
condition $\left\{ \rho_{\psi^\dagger},\chi_{\psi^\dagger}\right\}
= \left\{ \rho_\psi,\chi_\psi \right\} = 0$.) In the presence of
second-class constraints, consistent quantization requires that
the quantum anticommutator of two fields be identified with their
Dirac bracket (which depends on the Poisson bracket algebra of the
constraints) rather than with their classical Poisson bracket. To
the order of interest to us, the net effect of all this can be
implemented by saying that the following nonlinear field
redefinition restores the conventional anticommutation relations
and Fourier mode expansion (\ref{fmodexp}):
\begin{eqnarray}
\label{fldredef}
\tilde \rho_\alpha & = & \rho_\alpha \nn \\ \tilde \psi_\beta & =
& \psi_\beta
        +\frac{i}{8{}\Rhat^2}\biggl\{ (\psi'\psi)\psi_\beta
        - 2(\rho\Pi\psi)\psi_\beta - (\rho'\rho)\psi_\beta
        + 2(p_A {x'}^A)\rho_\beta \nn\\
& &     + \left[ (\rho'\psi)\rho_\beta
    - (\rho\psi')\rho_\beta\right] +2i{}\left[ y^2\psi_\beta
        + \frac{1}{2}\left( (p_A)^2
        + ({x'}^A)^2 \right)\psi_\beta\right]
        \biggr\}~ .
\end{eqnarray}
To be precise, we have removed the
second-class constraints on the fermionic variables by imposing,
via field redefinition, the proper quantization condition. In turn,
the Hamiltonian
has been recast in terms of fermi variables ($\rho_\psi$ and $\psi$) that
exhibit the usual anticommutation relations.  The
field redefinition has no effect on ${\cal H}_{int}$ to the order
of interest, but, when applied to ${\cal H}_{pp}$, it generates
new interaction terms of $O(1/\Rhat^2)$. This procedure gives rise
to non-trivial cancellations among terms that would potentially
renormalize the spectrum of supergravity modes. Without these
cancellations, the resulting spectrum of energy corrections would
be nonsensical.


Since we are treating these curvature corrections to the pp-wave
background in first-order perturbation theory, we are only
interested in physical string states that are eigenstates of the
pp-wave theory.  The level matching condition on these states is
met by fixing $x^{\prime -}$ such that $T_{01}$ vanishes at
leading order.  ($T_{01}$ is the current associated with
translation symmetry on the closed-string worldsheet.  Fixing
$T_{01}=0$ gives the usual level-matching condition for physical
pp-wave eigenstates.)

Conformal invariance demands that $T_{01}$ vanish order by order
in the expansion, and this is satisfied by fixing higher-order
corrections to $x^{\prime -}$. For some set of physical
eigenstates of the $1/\Rhat^2$ corrected geometry, the vanishing
of $T_{01}$ to this order would translate to an exact level
matching condition on these states.  Since we are not trying to
solve this theory exactly, this has no bearing on the present
calculation, and an explicit expression for $x^{\prime -}$ at
${\cal O}(1/\Rhat^2)$ is not needed.


The final result for the Hamiltonian density in the perturbed theory is
\begin{eqnarray}
\label{GSHamFinal}
{\cal H}={\cal H}_{pp}+{\cal H}_{int}\ , \qquad
{\cal H}_{int}={\cal H}_{BB}+{\cal
H}_{FF}+{\cal H}_{BF}~,
\end{eqnarray}
where
\begin{eqnarray}
{\cal H}_{pp} & = &
    \frac{1}{2}\left[(x^A)^2 + (p_A)^2 + ({x'}^A)^2\right]
    + \frac{i}{2}\left[
     \psi\psi' -  \rho \rho' + 2\rho\Pi\psi \right] ,
\end{eqnarray}
\begin{eqnarray}
\label{HBBfinal}
{\cal H}_{BB} & = & \frac{1}{\Rhat^2}\biggl\{
    \frac{1}{4}\left[ z^2\left( p_{y}^2 + {y'}^2 + 2{z'}^2 \right)
    -y^2\left( p_z^2 + {z'}^2 + 2{y'}^2\right)\right]
    + \frac{1}{8}\left[ (x^A)^2 \right]^2
\nn\\
& &     - \frac{1}{8}\left\{
    \left[ (p_A)^2\right]^2 + 2(p_A)^2({x'}^A)^2
    + \left[ ({x'}^A)^2\right]^2 \right\}
     + \frac{1}{2}\left({x'}^A p_A\right)^2
    \biggr\} ,
\end{eqnarray}
\begin{eqnarray}
{\cal H}_{\rm FF} & = &
        -\frac{1}{4 \Rhat^2}\biggl\{
  \left[ (\psi'\psi) + (\rho\rho')\right](\rho\Pi\psi)
        -\frac{1}{2}(\psi'\psi)^2 - \frac{1}{2}(\rho'\rho)^2
        + (\psi'\psi)(\rho'\rho)
\nn\\
& &     + (\rho\psi')(\rho'\psi)
    -\frac{1}{2}\left[ (\psi\rho')(\psi\rho') + (\psi'\rho)^2\right]
        + \biggl[
    \frac{1}{12}(\psi\gamma^{jk}\rho)(\rho\gamma^{jk}\Pi\rho')
\nn\\ & &     -\frac{1}{48}
        \left(\psi\gamma^{jk}\psi - \rho\gamma^{jk}\rho\right)
        \left(\rho'\gamma^{jk}\Pi\psi - \rho\gamma^{jk}\Pi\psi'\right)
        - (j,k \rightleftharpoons j',k') \biggr] \biggr\} ,
\end{eqnarray}
\begin{eqnarray}
\label{HBFfinal}
{\cal H}_{\rm BF} & = &
     \frac{1}{\Rhat^2}\biggl\{ -\frac{i}{4}\left[(p_A)^2+({x'}^A)^2
    + (y^2 - z^2)\right]\left(\psi\psi'- \rho\rho'\right)
\nn\\ & &     -\frac{1}{2}(p_A{x'}^A)(\rho\psi' + \psi\rho' )
        -\frac{i}{2}\left( p_k^2 + {y'}^2 -  z^2 \right)\rho\Pi\psi
\nn\\
& &     +\frac{i}{4}(z'_j z_k)\left(\psi\gamma^{jk}\psi -
    \rho\gamma^{jk}\rho\right)
        -\frac{i}{4}(y'_{j'} y_{k'})\left(\psi\gamma^{j'k'}\psi -
   \rho\gamma^{j'k'}\rho\right)
\nn\\
& &     -\frac{i}{8}(z'_k y_{k'} + z_k y'_{k'})
       \left(\psi\gamma^{kk'}\psi - \rho\gamma^{kk'}\rho\right)
        +\frac{1}{4{}}(p_k y_{k'} +  z_k p_{k'} )\psi\gamma^{kk'}\rho
\nn\\ & &     +\frac{1}{4{}}(p_j z'_k)\left(\psi\gamma^{jk}\Pi\psi
                + \rho\gamma^{jk}\Pi\rho\right)
        -\frac{1}{4{}}(p_{j'} y'_{k'})\left(\psi\gamma^{j'k'}\Pi\psi
                + \rho\gamma^{j'k'}\Pi\rho\right)
\nn\\ & &     
 -\frac{i}{2}(p_kp_{k'} - z'_k y'_{k'})\psi\gamma^{kk'}\Pi\rho  \biggr\} .
\nn\\
& &
\end{eqnarray}
Repeated indices are summed over; indices $j,k$ run over $1,..,4$ while
indices $i',j'$ run over $5,..,8$. The need for this notation arises
because the coordinates that descend from $AdS_5$ are treated differently
from those that descend from $S^5$. Put another way, the residual symmetry
of the problem is $SO(4)\times SO(4)$, not $SO(8)$.


\section{Quantization and Diagonalization of the Perturbation Hamiltonian}

The Hamiltonian (\ref{GSHamFinal}) is written using the same
conventions as in our discussion of the pp-wave limit in the first
section. To quantize it, we replace the canonical fields and
momenta by their expansion in string mode creation and
annihilation operators. The canonical commutation relations are
unchanged by the interactions, and we may therefore use the mode
expansions (\ref{modexpn},\ref{candaops}) and mode
(anti)commutators of the pp-wave limit without modification. The
terms quadratic in mode operators of course reproduce the pp-wave
Hamiltonian (\ref{ppham}). Terms quartic in mode oscillators
constitute the interaction Hamiltonian, the operator we must
diagonalize to find the perturbed spectrum. In this paper, we will
implement perturbation theory on the degenerate multiplets of
states created by acting on the ground state with two creation
operators. For these purposes, we only need the terms in $H_{int}$
containing two creation and two annihilation operators. As an
example of the outcome of this procedure, we display the purely
bosonic part of the expansion of $H_{int}$:
\begin{eqnarray}
\label{Hcorrected}
H_{BB} & = &
    -\frac{1}{32 J}\sum \frac{\delta(n+m+l+p)}{\xi}
    \times 
\nn\\
& & \biggl\{
    2 \biggl[ \xi^2 
	- (1 - k_l k_p k_n k_m )
     +  \omega_n \omega_m k_l k_p
      +  \omega_l \omega_p k_n k_m
    + 2 \omega_n \omega_l k_m k_p
\nn\\
& &     + 2 \omega_m \omega_p k_n k_l
    \biggr]
    a_{-n}^{\dagger A}a_{-m}^{\dagger A}a_l^B a_p^B
   +4 \biggl[ \xi^2 
	- (1 - k_l k_p k_n k_m )
     - 2 \omega_n \omega_m k_l k_p
     +  \omega_l \omega_m k_n k_p
\nn\\
& &   -  \omega_n \omega_l k_m k_p
    -  \omega_m \omega_p k_n k_l
    + \omega_n \omega_p k_m k_l \biggr]
    a_{-n}^{\dagger A}a_{-l}^{\dagger B}a_m^A a_p^B
     + 4  \biggl[8 k_l k_p
    a_{-n}^{\dagger i}a_{-l}^{\dagger j}a_m^i a_p^j
\nn\\
& &     + 2 (k_l k_p +k_n k_m)  
	a_{-n}^{\dagger i}a_{-m}^{\dagger i}a_l^j a_p^j
    +(\omega_l \omega_p+ k_l k_p -\omega_n 
	\omega_m- k_n k_m)a_{-n}^{\dagger i}a_{-m}^{\dagger i}a_l^{j'} a_p^{j'}
\nn\\
& &     -4 ( \omega_l \omega_p- k_l k_p)
	a_{-n}^{\dagger i}a_{-l}^{\dagger j'}a_m^i a_p^{j'} 
	-(i,j \rightleftharpoons i',j')
    \biggr]\biggr\} ,
\end{eqnarray}
where $\xi \equiv \sqrt{\omega_n \omega_m \omega_l \omega_p}$,
$\omega_n=\sqrt{1+k_n^2}$ and $k_n^2=\lambda^\prime n^2$. The
indices $l,m,n,p$ run from $-\infty$ to $+\infty$. The notation
distinguishes sums over indices in the first $SO(4)$ ($i,j,..$),
the second $SO(4)$ ($i^\prime,j^\prime,..$) and over the full
$SO(8)$ ($A,B,..$). Note that the powers of ${P_-}$ that come from
expanding the fields in terms of creation operators and doing the
integral over $\sigma$ combine to convert the small parameter
governing the strength of the perturbation from $1/\Rhat^2$ to
$1/J$, where $J$ is the integrally quantized (and large) angular
momentum of the string in the $S^5$ subspace. We have written
operator monomials in normal-ordered form.  Since ${\cal H}_{int}$
was derived as a classical object, the `correct' ordering of the
operators is not defined; we will allow for this ambiguity via an
appropriate normal-ordering constant. The corresponding oscillator
expressions for $H_{FF}$ and $H_{BF}$ are too complicated to
display here. 

To compute the string energy shifts induced by $H_{int}$ in first-order 
degenerate perturbation theory, we must find its matrix elements 
between the states of a multiplet degenerate under $H_{pp}$, and then
diagonalize the resulting finite-dimensional matrix. We will
execute this program on the 256-dimensional space of string
excited states created by acting on the ground state of angular
momentum $J$ with two creation operators of equal and opposite
moding (the latter condition is needed to satisfy the
level-matching constraint). The bosonic creation operators are
denoted ${a^A_n}^\dagger$, where $n$ is the integer mode index and
$A=1,\ldots,8$ is an $SO(8)$ vector index which decomposes as
$({\mathbf 4},{\mathbf 1})+ ({\mathbf 1},{\mathbf 4})$ under the
manifest $SO(4)\times SO(4)$ symmetry. Passing to a $SU(2)^2
\times SU(2)^2$ notation, we rewrite these representations as
$({\mathbf 2},{\mathbf 2};{\mathbf 1},{\mathbf 1}) + ({\mathbf
1},{\mathbf 1};{\mathbf 2},{\mathbf 2})$. The fermi creation
operators are ${b^\alpha_n}^\dagger$ where $\alpha$ is an $SO(8)$
spinor index which decomposes as $({\mathbf 2},{\mathbf
1};{\mathbf 2},{\mathbf 1}) + ({\mathbf 1},{\mathbf 2};{\mathbf
1},{\mathbf 2})$ under the spinor version of the manifest
$SO(4)\times SO(4)$ symmetry. The two four-dimensional irreps of
$SO(4)\times SO(4)$ are distinguished by their $\Pi$ eigenvalues
($+1$ or $-1$), and we will sometimes use $\Pi$ to categorize the
fermi creation operators accordingly. The eight bosonic and eight
fermionic creation operators allow us to create 256 `two-impurity'
states as follows:
\begin{eqnarray}
a_n^{A\dagger}
a_{-n}^{B\dagger} \ket{J} \qquad b_{n}^{\alpha\dagger}
        b_{-n}^{\beta\dagger}\ket{J}
\qquad a_n^{A\dagger} b_{-n}^{\alpha\dagger}\ket{J}
\qquad a_{-n}^{A\dagger} b_{n}^{\alpha\dagger}\ket{J}\ .
\end{eqnarray}
Half of these states are bosons and half are fermions. They
all have the same lightcone energy
\begin{eqnarray}
\Delta=2\sqrt{1+k_n^2}\sim 2+\lambda^\prime n^2 + \dots .
\end{eqnarray}
under $H_{pp}$. We expect to find non-zero matrix elements of
$H_{int}$ between these states according to the scheme shown in
table~\ref{blockform}. The matrix is block diagonal because half
the states in the multiplet are bosons and half are fermions, and
there are of course no matrix elements between the two. Because of
the complicated form of $H_{int}$ itself, especially in its
dependence on fermi fields, we found it necessary to use symbolic
manipulation programs to organize the calculation of explicit
forms for the matrices according to the various blocks in the
table. The results, as we will now show, turn out to be
surprisingly simple.
\begin{table}[ht!]
\begin{eqnarray}
\begin{array}{|c|cccc|}
\hline
 ({H})_{int} & a^{A\dagger}_n a^{B\dagger}_{-n} \ket{J} &
        b^{\alpha\dagger}_n b^{\beta\dagger}_{-n}\ket{J} &
        a^{A\dagger}_n b^{\alpha\dagger}_{-n} \ket{J} &
        a^{A\dagger}_{-n} b^{\alpha\dagger}_{n} \ket{J} \\
        \hline
\bra{J} a^{A}_n a^{B}_{-n} & { H}_{\rm BB} & { H}_{\rm BF} &0&0 \\
\bra{J} b^{\alpha}_n b^{\beta}_{-n} & { H}_{\rm BF} & { H}_{\rm
FF}&0&0\\ \bra{J} a^{A}_n b^{\alpha}_{-n} &0&0& { H}_{\rm BF} & {
H}_{\rm BF} \\ \bra{J} a^{A}_{-n} b^{\alpha}_n & 0 & 0 & { H}_{\rm
BF} & { H}_{\rm BF}\\
\hline
\end{array} \nonumber
\end{eqnarray}
\caption{Structure of the matrix of first-order energy
perturbations in the space of two-impurity string states}
\label{blockform}
\end{table}

The matrix elements of $H_{int}$ between spacetime bosons built out of
bosonic string oscillators only turn out to have the following explicit form:
\begin{eqnarray}
\label{bosonmatrix}
 \Braket{ J | a_n^A a_{-n}^B \left( {H}_{BB} \right)
        a_{-n}^{C \dagger} a_n^{D \dagger} | J }   & = & 
        \left( N_{BB}(k_n^2) - 2 n^2\lambda'\right)
	\frac{\delta^{ AD}\delta^{ BC}}{J}
\qquad\qquad\qquad\nn\\
&&  	+ \frac{n^2\lambda'}{J(1+n^2\lambda')}      
	\left[ \delta^{ab}\delta^{cd}
        + \delta^{ad}\delta^{bc} - \delta^{ac}\delta^{bd} \right]
\nn\\
&&	- \frac{n^2\lambda'}{J(1+n^2\lambda')}  	 
	\left[ \delta^{a'b'}\delta^{c'd'}
        + \delta^{a'd'}\delta^{b'c'} - \delta^{a'c'}\delta^{b'd'} \right]
\nn\\
 \approx  \left(n_{BB}-2 \right)&&\kern-25pt   
	\frac{n^2\lambda'}{J} \delta^{ AD}\delta^{ BC}
	    + \frac{n^2\lambda'}{J}\left[ \delta^{ab}\delta^{cd}
        + \delta^{ad}\delta^{bc} - \delta^{ac}\delta^{bd} \right]
\nn\\
& &     - \frac{n^2\lambda'}{J}\left[ \delta^{a'b'}\delta^{c'd'}
        + \delta^{a'd'}\delta^{b'c'} - \delta^{a'c'}\delta^{b'd'} \right]
	+ {O}({\lambda'}^{2})\ ,
\end{eqnarray}
where lower-case $SO(4)$ indices $a,b,c,d\in 1,\dots ,4$ mean
that the corresponding $SO(8)$ labels $A,B,C,D$ all lie in the
first $SO(4)$, while the indices $a',b',c',d'\in 5,\dots ,8$ mean
that the $SO(8)$ labels lie in the second $SO(4)$ $(A,B,C,D \in
5,\dots ,8)$. Note that we have written both the exact matrix element
and its expansion in powers of $\lambda'$. Since $\lambda'$ is related
to the gauge coupling constant by the AdS/CFT correspondence, the
$\lambda'$ expansion of the energy eigenvalues is what is needed to
make comparisons with perturbative gauge theory calculations of 
operator dimensions. 

A further important point is that we have 
included a function $N_{BB}(k_n^2)$ to account for operator ordering 
ambiguities. In a matrix element of $H_{int}$ between states of this 
type, different operator orderings differ by terms proportional 
to $\delta^{AD}\delta^{BC}$ (the ambiguity arises from commuting 
creation and annihilation operators of the same mode number past each 
other: $A,D$ go with mode $n$ and $B,C$ go with mode $-n$). The 
coefficient of this term is an a priori arbitrary function of 
$k_n^2=n^2\lambda^\prime$ which we will define by its power series.
In order for the energy shift to be perturbative in the gauge coupling 
(i.e.~to vanish as $\lambda'\to 0$) the $k_n^0$ term in the power series 
must vanish. Therefore we can write $N_{BB}(k_n^2)=n_{BB}k_n^2+O(k_n^4)$,
which says that the one-loop ($O(\lambda')$ or $O(k_n^2)$) normal-ordering 
ambiguity is contained in the single constant $n_{BB}$.  
Although we are careful to include such additions at this level,
it will be shown that these extra normal-ordering 
constants must vanish to all orders; the standard operator-ordering 
prescription used to define (\ref{Hcorrected}) is correct as it stands.

The matrix elements between bosonic states created by pairs of fermionic
creation operators have a remarkably similar form:
\begin{eqnarray}
\label{fermimatrix}
\Braket{J| b_n^\alpha b_{-n}^\beta\left({H}_{FF}\right)
          b_{-n}^{\gamma\dagger} b_n^{\delta\dagger}|J} & = &
    \left(N_{FF}(k_n^2)-2 {n^2\lambda'}\right) 
	\frac{\delta^{\alpha\delta}\delta^{\beta\gamma}}{J} \qquad\qquad\qquad\nn\\
     + \frac{n^2\lambda'}{24 J(1+n^2\lambda')} &&\kern-25pt
\left[ (\gamma^{ij})^{\alpha\delta}(\gamma^{ij})^{\beta\gamma}
        + (\gamma^{ij})^{\alpha\beta}(\gamma^{ij})^{\gamma\delta}
        - (\gamma^{ij})^{\alpha\gamma}(\gamma^{ij})^{\beta\delta} \right]\nn\\
    - \frac{n^2\lambda'}{24 J(1+n^2\lambda')} &&\kern-25pt
    \left[(\gamma^{i'j'})^{\alpha\delta}(\gamma^{i'j'})^{\beta\gamma}
        + (\gamma^{i'j'})^{\alpha\beta}(\gamma^{i'j'})^{\gamma\delta}
-
(\gamma^{i'j'})^{\alpha\gamma}(\gamma^{i'j'})^{\beta\delta}\right]
\nn\\
\approx	\left( n_{FF}-2 \right) 
	&&\kern-25pt\frac{n^2\lambda'}{J}\delta^{\alpha\delta}\delta^{\beta\gamma} 
     	+ \frac{n^2\lambda'}{24 J} 
\left[ (\gamma^{ij})^{\alpha\delta}(\gamma^{ij})^{\beta\gamma}
        + (\gamma^{ij})^{\alpha\beta}(\gamma^{ij})^{\gamma\delta}
        - (\gamma^{ij})^{\alpha\gamma}(\gamma^{ij})^{\beta\delta} \right]
\nn\\
    - \frac{n^2\lambda'}{24 J} &&\kern-25pt
    \left[(\gamma^{i'j'})^{\alpha\delta}(\gamma^{i'j'})^{\beta\gamma}
        + (\gamma^{i'j'})^{\alpha\beta}(\gamma^{i'j'})^{\gamma\delta}
-
(\gamma^{i'j'})^{\alpha\gamma}(\gamma^{i'j'})^{\beta\delta}\right]
	+ {O}({\lambda'}^2)~.
\nn\\ & & 
\end{eqnarray}
The discussion of the normal-ordering function $N_{FF}$ follows exactly 
the same lines as the discussion of $N_{BB}$ in (\ref{bosonmatrix}). The
gamma matrices are $SO(8)$ generators, lower-case Roman characters
are $SO(8)$ indices with the prime/unprime notation distinguishing
$i\in 1,2,3,4$ from $i'\in 5,6,7,8$. Repeated indices are summed
over. Note that the generators $\gamma^{ij}$ and $\gamma^{i'j'}$
all act within one or the other $SO(4)$, and therefore commute
with $\Pi=\gamma^1\gamma^2\gamma^3\gamma^4$. A careful analysis
shows that $H_{FF}$ is non-zero only for transitions of the types
$++\to ++$ and $--\to --$ (using $\pm$ to denote the $\Pi$
eigenvalues of the two fermionic mode operators).

The matrix that mixes bi-fermionic bosons with ordinary bosons
has the interesting structure
\begin{eqnarray}
\label{bfmix}
\Braket{J| b_{n}^\alpha b_{-n}^\beta \left( {H}_{BF} \right)
        a_{-n}^{A\dagger} a_{n}^{B\dagger} |J} & = &
	\frac{n^2 {\lambda'}}{2J(1+n^2\lambda')}
	\biggl\{
	\sqrt{1+n^2\lambda'}\Bigl[
                \left( \gamma^{ab'} \right)^{\alpha\beta}
                - \left( \gamma^{a'b} \right)^{\alpha\beta} \Bigr]
\nn\\
& & 	+~ n\sqrt{ \lambda' }\left[
	\left( \gamma^{a'b'} \right)^{\alpha\beta}
	- \left( \gamma^{ab} \right)^{\alpha\beta}
	+ \left(\delta^{ab} - \delta^{a'b'}\right)
	\delta^{\alpha\beta} \right]
	\biggr\}
\nn\\
& \approx &
	\frac{ n^2 \lambda'}{2 J}\left[
                \left( \gamma^{ab'} \right)^{\alpha\beta}
                - \left( \gamma^{a'b} \right)^{\alpha\beta} \right]
	+{O}({\lambda'}^{3/2})~. 
\end{eqnarray}
The complex conjugate of this gives the other off-diagonal component
of the bosonic block of the perturbation matrix of $H_{int}$. Note that
there is no operator-ordering ambiguity in this matrix element and thus no
need for an adjustable normal-ordering constant. The appearance of
factors of $\sqrt{\lambda'}$ in the matrix elements is alarming, since
it could lead to string energies that are not analytic at the origin  
and which could therefore not be matched to gauge perturbation theory.
Fortunately, to the order we have explored, the half-integer powers of
$\lambda'$ cancel out of the string energies so that the spectrum is 
in fact analytic in $\lambda'$. We do not have a general proof
of this important property at the moment. For future use, we note
that the leading order in $\lambda'$ limit of the matrix element
(the last line in (\ref{bfmix})) vanishes unless the two bosonic mode 
operators are in different $SO(4)$'s. The limiting matrix element also
vanishes unless the two fermionic mode operators have opposite $\Pi$ values 
(this is because a $\gamma^{ab'}$ has one gamma matrix from each $SO(4)$ and 
anticommutes with the matrix $\Pi=\gamma^1\gamma^2\gamma^3\gamma^4$). 

Finally, we record the matrix elements of $H_{BF}$ between fermionic states 
created by acting on the string ground state with one bosonic and one 
fermionic creation operator (the lower block of the perturbation matrix):
\begin{eqnarray}
\label{22}
\Braket{ J | b_{n}^\alpha a_{-n}^A \left( {H}_{BF} \right)
    b_{n}^{\beta\dagger} a_{-n}^{B\dagger}|J } & = &
	N_{BF}(k_n^2)\frac{\delta^{AB}\delta^{\alpha\beta}}{J}
\nn\\
 	+ \frac{n^2\lambda'}{2J(1+n^2\lambda')}\biggl\{&&\kern-30pt
	\left( \gamma^{ab}\right)^{\alpha\beta}
        - \left( \gamma^{a'b'}\right)^{\alpha\beta} 
	- (3+4n^2\lambda') \delta^{ab} \delta^{\alpha\beta}
	- (5+4n^2\lambda')\delta^{a'b'}\delta^{\alpha\beta}
	\biggr\}
\nn\\
   \approx 
	 \frac{n^2 \lambda'}{2J}
	\biggl\{\left( \gamma^{ab}\right)^{\alpha\beta} 
        - \Bigl( &&\kern-30pt  \gamma^{a'b'}  \Bigr)^{\alpha\beta}
    + \left[ (2 n_{BF}-3) \delta^{ab}+ (2 n_{BF}-5) \delta^{a'b'}
	 \right]\delta^{\alpha\beta} \biggr\} 
	+ {O}({\lambda'}^{2})~,
\nn\\
& & 
\end{eqnarray}
\begin{eqnarray}
\label{22a}
\Braket{J | b_{n}^\alpha a_{-n}^A \left({H}_{BF}\right)
        b_{-n}^{\beta\dagger} a_{n}^{B\dagger}|J } & = &
\frac{n^2\lambda'}{2J\sqrt{1+n^2\lambda'}}\biggl\{
	\left( \gamma^{ab}\right)^{\alpha\beta}
        - \left( \gamma^{a'b'}\right)^{\alpha\beta} 
\nn\\
 	-\frac{n {\lambda'}^{1/2}}{\sqrt{1+n^2\lambda'}}
	&&\kern-25pt\Bigl[
	\left( \gamma^{ab'}\right)^{\alpha\beta}
        - \left( \gamma^{a'b}\right)^{\alpha\beta} \Bigr]
	-\delta^{\alpha\beta}
	\left( \delta^{ab} - \delta^{a'b'} \right)
	\biggr\}
\nn\\
  \approx  \frac{n^2 \lambda'}{2 J}
	\biggl\{
         \left( \gamma^{ab}\right)^{\alpha\beta}&&\kern-25pt
        -  \left( \gamma^{a'b'}\right)^{\alpha\beta}
    - \left( \delta^{ab}- \delta^{a'b'} \right)\delta^{\alpha\beta} \biggr\}
	+ {O}({\lambda^\prime}^{3/2})\ .
\end{eqnarray}
The matrix elements have terms that are non-analytic in $\lambda'$, but it
once again turns out that the energy eigenvalues are analytic in $\lambda'$ 
(as must be the case to make contact with perturbative gauge theory).

Equation (\ref{22}) has its own normal-ordering function, 
$N_{BF}$, but the structure of the perturbing Hamiltonian implies that 
it is related to the other normal-ordering functions by 
$N_{BF} = N_{BB} + N_{FF}$ \cite{CGCIanTristan}. 
It turns out that $N_{BB}$ alone shifts the energies of string
states that correspond to the dimensions of operators at 
supermultiplet levels $L=0,8$.  
For finite $\lambda'$, these levels are 
shifted by the function $N_{BB}$ itself; in the small-$\lambda'$ expansion 
they are shifted by some constant coefficient at each order in the series.
In the same way, $N_{FF}$ and $N_{BB}$ provide energy shifts to levels
$L=2,4,6$, and $N_{BF}$ controls $L=1,3,5,7$.   
In the gauge theory,
supersymmetry dictates that the level spacing must be uniform throughout 
the supermultiplet, i.e.~the spectrum of anomalous dimensions is a linear function 
of $L$ (\ref{deltalevel}).
To meet this condition in the string theory, we require 
$N_{BB} = N_{BF}$.  Furthermore, levels $L=2,4,6$ are populated in such a 
way that $N_{BB}$ must also be equal to $N_{FF}$.
Combined with the above constraint $N_{BF} = N_{BB} + N_{FF}$,
however, the normal-ordering functions 
must vanish to all orders in $\lambda'$.  We therefore set
$N_{BB} = N_{FF} = N_{BF} = 0$, which eliminates
all normal-ordering ambiguity from the string theory.

An additional observation about all of the above matrix elements is that 
they vanish for $n=0$. This means that states made from two zero-mode 
oscillators receive no interaction corrections (because of the 
level-matching constraint, these are the only two-impurity states
that involve zero modes). As has been argued elsewhere in the paper, 
non-renormalization of the zero-mode oscillators is the simplest 
way to understand how the string states manage to reproduce the 
large degeneracies implicit in the $PSU(2,2|4)$ superconformal 
symmetry. To put this conjecture to a more stringent test, we 
would have to look at higher-impurity states, an exercise we will
defer to a subsequent paper. The calculations done here just scratch 
the surface of this subject, but are at least consistent with the 
larger conjecture. 

We now turn to the problem of finding the eigenvalues of $H_{int}$ and
comparing the results with gauge theory predictions. Given the functional
form of the matrix elements of the string theory perturbing Hamiltonian 
(\ref{bosonmatrix}-\ref{22a}), the energy eigenvalues will in general be 
fairly complicated functions of $\lambda'$. However, since we want to 
compare them to perturbative gauge theory anomalous dimensions (which
are found as power expansions in $\lambda'$), we can simplify the analysis 
by expanding the string Hamiltonian matrix elements to the appropriate 
order in $\lambda'$ before diagonalizing. Since it is instructive, we will 
first do the leading-order calculation of the string spectrum and its 
comparison with one-loop anomalous dimensions in some detail. 
We will then quote the results of the calculation of the string spectrum 
to higher orders in $\lambda'$ and discuss their comparisons with 
recently-determined higher-loop gauge theory anomalous dimensions 
\cite{Beisert:2003tq}. 

If we expand the string Hamiltonian $H=H_{pp}+H_{int}$ to first order in 
$\lambda'$, we know that the energy eigenvalues will have the general form
\begin{equation}
E(n,J) =  2 + \lambda^\prime n^2
        \left( 1 + \frac{\Lambda}{J} + {O}({J}^{-2}) \right)\ ,
\end{equation}
where $\Lambda$ is a dimensionless quantity that distinguishes the
different eigenvalues of $H_{int}$. This is to be compared with the
generic formula for one-loop anomalous dimensions of gauge theory 
operators of large $R$-charge (\ref{finlopdim}):
\begin{equation}
\Delta(n,R) =  2 + \frac{g^2_{YM} N_c}{R^2} n^2
        \left( 1 + \frac{\bar\Lambda}{R} + {O}({R}^{-2}) \right)\ ,
\end{equation}
where, in this case, $\bar\Lambda$ is a dimensionless quantity that depends
on the operator multiplet. The AdS/CFT correspondence asserts that, with
the identifications $R\simeq J$ and $g^2_{YM} N_c/R^2\simeq\lambda^\prime$,
the two expressions should match. This will indeed be true, provided
that the two ways of calculating $\Lambda$ give the same result.

The manifest $SO(4)_{AdS}\times SO(4)_{S^5}$ transverse space
symmetry of the problem can be used to classify eigenvectors and
greatly simplify the diagonalization problem in the one-loop limit. 
We begin with the 128-dimensional block of table \ref{blockform} 
that acts on bosonic two-impurity states. Recall that the 
leading-order-in-$\lambda'$ matrix elements of the mixing
Hamiltonian $H_{BF}$ (\ref{bfmix}) vanish on states created by two
bosonic oscillators from the same $SO(4)$, and also on states
created by two fermionic oscillators with the same $\Pi$
eigenvalue.  Thus, $H_{BB}$ and $H_{FF}$ may be independently
diagonalized (without worrying about boson-fermion mixing) on
these two separate 32-dimensional subspaces. $H_{BF}$ has
non-vanishing matrix elements on the orthogonal 64-dimensional
subspace spanned by two bosonic creation operators from different
$SO(4)$'s and two fermionic creation operators of opposite $\Pi$
eigenvalue: it poses a separate diagonalization problem which
mixes boson-boson with fermion-fermion states.

\begin{table}[ht!]
\begin{equation}
\begin{array}{|c|c|}\hline
 SO(4)_{AdS}\times  SO(4)_{S^5} & \Lambda_{BB} \\
\hline ({\bf 1,1;1, 1}) &  {-6  } \\ ({\bf 1,1; 3,3})
&{-2 } \\ ({\bf 1,1;3,1}) +({\bf 1,1;1,3})& {-4  } \\ \hline
  ({\bf 1,1;1, 1}) & 2  \\
 ({\bf 3,3;1, 1}) & {-2 } \\
  ({\bf 3,1;1, 1}) + ({\bf 1,3 ;1, 1})& { 0 } \\
\hline
\end{array} \nonumber
\end{equation}
\caption{$O(1/J)$ energy shifts for various bosonic modes}
\label{bosonspectrum}
\end{table}

We start with the diagonalization of $H_{BB}$ on states created by
two bosonic mode oscillators in the same $SO(4)\approx SU(2)\times
SU(2)$. In $SU(2)$ notation, a four-vector of $SO(4)$ is
represented as $({\bf 2,2})$. Using this notation, the bosonic
modes are in the $SO(4) \times SO(4)$ representations $({\bf
2,2;1, 1})+ ({\bf 1,1;2,2})$, and the representation content of
the states created by two such oscillators is given by the $SO(4)$
formula $({\bf 2,2})\times ({\bf 2,2}) = ({\bf 3,3})+({\bf
3,1})+({\bf 1, 3})+({\bf 1,1})$. ${H}_{BB}$ is diagonalized by
simply projecting it onto the different invariant subspaces. We
find the values for $\Lambda_{\rm BB}$ shown in
table~\ref{bosonspectrum}. The total number of states in the table
is 32. The remaining 32 states of the form ${a_n^a}^\dagger
{a_{-n}^{b'}}^\dagger\ket{J}$ and ${a_n^{a'}}^\dagger
{a_{-n}^{b}}^\dagger\ket{J}$ are subject to bose-fermi mixing and
will be dealt with shortly.

\begin{table}[ht!]
\begin{eqnarray}
\begin{array}{|c|c|c|c|}\hline
{\rm Operator} &  SO(4)_{AdS}\times SO(4)_{S^5}& \Delta &\bar\Lambda\\
\hline
\Sigma_A\tr\left(\phi^AZ^p\phi^AZ^{R-p}\right)
 &({\bf 1,1;1, 1}) & 2+\frac{g^2_{YM} N_c}{R^2}~n^2(1-\frac{6}{R})& -6 \\
\tr\left(\phi^{(i}Z^p\phi^{j)}Z^{R-p}\right)
 &({\bf 1,1; 3,3}) & 2+\frac{g^2_{YM} N_c}{R^2}~n^2(1-\frac{2}{R}) & -2\\
\tr\left(\phi^{[i}Z^p\phi^{j]}Z^{R-p}\right)
 &({\bf 1,1;3,1}) +({\bf 1,1;1,3})&2+\frac{g^2_{YM} N_c}{R^2}~n^2
(1-\frac{4}{R})&-4\\
\hline
\tr\left(\nabla_\mu Z Z^p\nabla^\mu Z Z^{R-2-p}\right)
 &({\bf 1,1;1, 1})& 2+\frac{g^2_{YM} N_c}{R^2}~n^2(1+\frac{2}{R}) & 2\\
\tr\left(\nabla_{(\mu}ZZ^p\nabla_{\nu)}Z Z^{R-2-p}\right) &
 ({\bf 3,3;1, 1}) & 2+\frac{g^2_{YM} N_c}{R^2}~n^2(1-\frac{2}{R}) & -2\\
\tr\left(\nabla_{[\mu}ZZ^p\nabla_{\nu]}Z Z^{R-2-p}\right) & ({\bf
3,1;1, 1}) + ({\bf 1,3 ;1, 1}) & 2+\frac{g^2_{YM}
N_c}{R^2}~n^2(1-\frac{0}{R}) & 0 \\ \hline
\end{array} \nonumber
\end{eqnarray}
\caption{Gauge operators corresponding to string theory
energy eigenstates listed in table~\ref{bosonspectrum}}\label{BBmatch}
\end{table}

We now want to ask whether these energy shifts match gauge theory
predictions. Tables~\ref{tableone} and \ref{tabletwo} in section 3
list predictions for the one-loop anomalous dimension of various
operators in a supermultiplet. Among them are a number of purely
bosonic operators, distinguished mainly by their different
transformation properties under an $SO(4)$ $R$-charge. Information
provided by the group theory analysis allows us to uniquely match
these operators to the string energy eigenstates listed in the
previous table, with the result shown in table~\ref{BBmatch}. In
the column labeled $\Delta$, we list the expansion of the
anomalous dimension as determined from the general formula
(\ref{deltalevel}) for the relevant supermultiplet. In the last
column we list the inferred value of $\bar\Lambda$.  The results
agree with the corresponding sector of the string spectrum.
\begin{table}[ht!]
\begin{equation}
\begin{array}{|c|c|}\hline
 SO(4)_{AdS}\times SO(4)_{S^5} & \Lambda_{FF} \\
\hline ({\bf 1},{\bf 1};{\bf 1},{\bf 1})& {-2 }
\\ ({\bf 1},{\bf 1};{\bf 3},{\bf 1})& {0} \\
({\bf 3},{\bf 1};{\bf 1},{\bf 1})& {-4 } \\ ({\bf
3},{\bf 1};{\bf 3},{\bf 1})& {-2 } \\ \hline
\end{array} \nonumber
\end{equation}
\caption{Energy shifts of states created by two fermionic
oscillators of the same $\Pi$ eigenvalue} \label{fermispectrum}
\end{table}

Now we turn to the diagonalization of $H_{FF}$ on the
32-dimensional subspace spanned by states created by two fermionic
oscillators of the same $\Pi$ eigenvalue. To classify the
fermions, we use that the $\Pi=+$ oscillators transform as $({\bf
2, 1};{\bf 2, 1})$ and the $\Pi=-$ oscillators transform as $({\bf
1, 2};{\bf 1, 2})$. The matrix elements (\ref{fermimatrix}) only
connect $++$ states with $++$ and $--$ with $--$ (and the two
submatrices are identical), so the problem of diagonalizing
$H_{FF}$ is further simplified. The results for $\Lambda_{FF}$,
obtained by projecting onto the different invariant subspaces, are
shown in table~\ref{fermispectrum}. There is another copy of this
spectrum in which the roles of the two $SU(2)$ factors inside each
$SO(4)$ are interchanged. These states can be unambiguously
matched to the $\Delta_0 = 2$ gauge theory operators built out of
pairs of gluino fields. The one-loop anomalous dimensions of
some of the operators of this type, along with the predicted values
of $\bar\Lambda$ are shown in table~\ref{FFmatch} (the notation
for the operator monomials is rather compressed, but self-explanatory
we hope).  The match with the corresponding string energy eigenvalues
and multiplicities is perfect.
\begin{table}[ht!]
\begin{eqnarray}
\begin{array}{|c|c|c|c|}\hline
{\rm Operator} & SO(4)_{AdS}\times SO(4)_{S^5}&
    \Delta &\bar\Lambda\\ \hline
\tr\left(\chi^{[\alpha}Z^p\chi^{\beta]}Z^{R-1-p}\right)
    &({\bf 1},{\bf 1};{\bf 1},{\bf 1})&
    2+\frac{g^2_{YM} N_c}{R^2}~n^2(1-\frac{2}{R})& -2 \\
\tr\left(\chi^{(\alpha}Z^p\chi^{\beta)}Z^{R-1-p}\right)
    &({\bf 1},{\bf 1};{\bf 3},{\bf 1})&
    2+\frac{g^2_{YM} N_c}{R^2}~n^2(1-\frac{0}{R})& 0 \\
\tr\left(\chi[\sigma_{\mu},\tilde\sigma_\nu] Z^p\chi Z^{R-1-p}\right)
    &({\bf 3},{\bf 1};{\bf 1},{\bf 1})&
    2+\frac{g^2_{YM} N_c}{R^2}~n^2(1-\frac{4}{R})& -4 \\
\hline
\end{array} \nonumber
\end{eqnarray}
\caption{Two gluino operators corresponding to some of the states
in table~\ref{fermispectrum}} \label{FFmatch}
\end{table}

To complete our discussion of the bosonic spectrum, we have to
diagonalize $H_{int}$ on the remaining 64-dimensional subspace of
states on which $H_{BF}$ has non-zero matrix elements. As
explained earlier in the discussion following (\ref{bfmix}),
the states on which $H_{BF}$ acts nontrivially at leading order 
in $\lambda'$, are created by two bosonic oscillators taken from
different $SO(4)$s, or by two fermionic oscillators with different 
$\Pi$ eigenvalues. For two such
fermions, the representation content is $({\bf 2, 1};{\bf 2,
1})\times ({\bf 1, 2};{\bf 1, 2})$, while for two bosons, it is
$({\bf 2, 2};{\bf 1, 1})\times ({\bf 1, 1};{\bf 2, 2})$. The net
result in both cases is $({\bf 2, 2};{\bf 2, 2})$ (or $({\bf 4,
4})$ in the $SO(4)$ language). There are two distinct ways of
distributing mode indices on the creation operators, and therefore
two realizations of this representation for each of the bosonic
and fermionic cases. The eigenvalue problem therefore reduces to
that of diagonalizing a $4\times 4$ numerical matrix. Note that
$H_{BB}$ and $H_{FF}$ also have matrix elements between these
states.
The results of the diagonalization, 
which are extremely simple, are given in table~\ref{mixspectrum}.

\begin{table}[ht!]
\begin{equation}
\begin{array}{|c|c|}\hline
 SO(4)_{AdS}\times SO(4)_{S^5} & \Lambda_{\rm BF} \\
\hline ({\bf 2},{\bf 2};{\bf 2},{\bf 2})& -4 \\ ({\bf 2},{\bf
2};{\bf 2},{\bf 2}) & -2 \\ ({\bf 2},{\bf 2};{\bf
2},{\bf 2})& 0  \\ \hline
\end{array} \nonumber
\end{equation}
\caption{String eigenstates in the subspace for which $H_{BF}$ has
non-zero matrix elements} \label{mixspectrum}
\end{table}

\begin{table}[ht!]
\begin{eqnarray}
\begin{array}{|c|c|c|c|}\hline
{\rm Operator} & SO(4)_{AdS}\times SO(4)_{S^5}&
        \Delta &\bar\Lambda\\ \hline
\tr\left(\phi^i Z^p \nabla_\mu Z Z^{R-1-p}\right) + \dots
        &({\bf 2},{\bf 2};{\bf 2},{\bf 2})&
        2+\frac{g^2_{YM} N_c}{R^2}~n^2(1-\frac{4}{R})& -4 \\ \hline
\tr\left(\phi^i Z^p \nabla_\mu Z Z^{R-1-p}\right)
        &({\bf 2},{\bf 2};{\bf 2},{\bf 2})&
        2+\frac{g^2_{YM} N_c}{R^2}~n^2(1-\frac{2}{R})& -2 \\ \hline
\tr\left(\phi^i Z^p \nabla_\mu Z Z^{R-1-p}\right) +\dots
        &({\bf 2},{\bf 2};{\bf 2},{\bf 2})&
        2+\frac{g^2_{YM} N_c}{R^2}~n^2(1-\frac{0}{R})& 0 \\ \hline
\end{array} \nonumber
\end{eqnarray}
\caption{Gauge theory operators corresponding to the string states
in table~\ref{mixspectrum}} \label{BFmatch}
\end{table}

The eigenvectors are simple linear combinations of the bosonic and
fermionic basis states. The corresponding operators and their
one-loop dimensions, given in table~\ref{BFmatch}, can be read off
from Beisert's listing of the elements of the gauge theory
superconformal multiplet. The ellipses indicate that the operator
in question contains further terms involving fermi fields. (See
App.~B of Beisert \cite{BeisertSUSY} for details.) The gauge
theory analysis says that there is operator mixing between bosonic
and fermionic operators of certain restricted symmetry properties and
these are precisely the operators that mix according to the string
theory analysis. The gauge theory analysis of boson-fermion mixing
is, on the face of it, very complicated, but a simple outcome is
legislated by the supersymmetry analysis. This outcome matches
perfectly the first-order perturbation theory of string energy
levels.

Carrying the above line of argument through to the end, one can
find the first order energy shifts of the 128 spacetime bosons of
angular momentum $J$ and count the degeneracies of the shifted
levels. The results are given in table~\ref{boseshifts}.
\begin{table}[ht!]
\begin{equation}
\begin{array}{|c|ccccc|}\hline
{\rm Level} & 0 & 2 & 4 & 6 & 8 \\
\hline
{\rm Mult.} & 1 & 28 & 70 & 28 & 1 \\
\hline
\Lambda_{bose}   & -6 & -4 & -2 & 0 & 2 \\ \hline
\end{array} \nonumber
\end{equation}
\caption{First-order energy shifts of the 128 spacetime bosons of
angular momentum $J$} \label{boseshifts}
\end{table}

We can also carry out the same exercise for the 128 spacetime fermions.
The results for energy shifts and multiplicities
are given in table~\ref{fermishifts}.
\begin{table}[ht!]
\begin{equation}
\begin{array}{|c|cccc|} \hline
{\rm Level} & 1 & 3 & 5 & 7 \\
\hline
{\rm Mult.} & 8 & 56 & 56 & 8  \\
\hline
\Lambda_{fermi}   & -5 & -3 & -1 & 1 \\ \hline
\end{array} \nonumber
\end{equation}
\caption{First-order energy shifts of the 128 spacetime fermions
of angular momentum $J$} \label{fermishifts}
\end{table}
These two energy/multiplicity tables, taken together, perfectly
reproduce the table of one-loop gauge theory anomalous dimension
predictions summarized in table~\ref{smultiplicity}. Although we
have not presented the complete analysis, the break-up of the
various degenerate submultiplets into $SO(4)\times SO(4)$ irreps
matches the gauge theory predictions as well.

Now, as mentioned earlier, a certain amount of higher-order information
about dimensions of the relevant operators has become available, and
this should allow us to carry out a new and independent set of
cross-checks between gauge theory and string theory.  To explore 
this issue, it is useful to have the exact, all-orders in
$\lambda'$ expression for the $O(1/J)$ shift in string energies.
In other words, we now want to diagonalize the matrix defined by 
(\ref{bosonmatrix}-\ref{22a}) without expanding in $\lambda'$.
This is slightly more painful than the one-loop diagonalization:
when we go beyond leading order in $\lambda'$, the bose-fermi 
interaction term $H_{BF}$ (\ref{bfmix}) mixes bosonic indices in 
both of the $SO(4)$ subgroups and this enlarges the effective
size of the matrices which must be diagonalized. Nevertheless, 
with the help of symbolic manipulation programs, one can get a 
formula for the string theory energy corrections for all the states 
in the supermultiplet, exact to all orders in $\lambda'$.  
The final result is the following concise formula for the energy 
levels including shifts of $O(1/J)$:
\begin{eqnarray}
\label{stringfinal}
E_L(n,J) & = & 2\sqrt{1+\lambda' n^2} 
	-\frac{n^2\lambda'}{J}\left[
	2+\frac{(4-L)}{\sqrt{1+n^2\lambda'}}\right]+O(1/J^2)~.
\end{eqnarray}
For comparison with the gauge theory, we expand in small $\lambda'$:
\begin{eqnarray}
\label{stringfinalexp}
E_L(n,J)  & \approx& \left[ 2 + \lambda' n^2 -
	 \frac{1}{4}(\lambda' n^2)^2 + \frac{1}{8}(\lambda' n^2)^3 
	+\dots \right]  \nonumber\\ 
	&~ & + \frac{1}{J}\left[ n^2\lambda'(L-6)+
	(n^2\lambda')^2\left(\frac{4-L}{2}\right)+
	(n^2\lambda')^3\left(\frac{3L-12}{8}\right) +\ldots\right]~.
\end{eqnarray}
The exact energy is organized into degenerate sub-levels $L=0,\ldots,8$ with 
the same multiplicities and $SO(4)\times SO(4)$ content as we found for 
the one-loop energy shift. The expansion in powers of $\lambda'$ shows
that the $\lambda'/J$ term matches, as it should, the one-loop results
summarized in tables \ref{boseshifts} and \ref{fermishifts}. The higher-order 
terms amount to predictions for two-loop and three-loop
(and higher-loop, if one wanted) gauge theory anomalous dimensions. 
Note that the shift is predicted to vanish beyond
first order for the $L=4$ level, the level which includes the bosonic 
$SU(4)$ irrep that is forbidden to mix with fermionic operators.

Parts of this prediction can in fact be checked: 
Beisert, Kristjansen and Staudacher \cite{Beisert:2003tq} 
have computed the two-loop correction to 
anomalous dimensions of certain BMN operators in the gauge theory
planar limit. The calculation is done for operators 
at level four in the supermultiplet, for which they find the 
following expression: 
\begin{eqnarray}
\label{twoloopgauge}
\delta\Delta_n^R & = & -\frac{g_{YM}^4 N_c^2}{\pi^4}
	\sin^4\frac{n\pi}{R+1}\left(
	\frac{1}{4}+\frac{\cos^2\frac{n\pi}{R+1}}{R+1}\right)~,
\end{eqnarray} 
where $R$ is the $R$-charge of the operator.
(The authors of \cite{Beisert:2003tq} use the symbol $J$ instead of $R$.)
The discussion leading up to (\ref{deltalevel}) implies that
${\cal N}=4$ supersymmetry allows us to infer the dimensions of all 
operators at all levels of this supermultiplet by making the substitution 
$R \to R+2-{L}/{2}$ in the formula for the dimension of the $L=4$ 
operator ($R$ is the R-charge of the operator, whatever the level): 
\begin{eqnarray}
\label{twoloopgaugeL}
\delta\Delta_n^{R,L} & = & -\frac{g_{YM}^4 N_c^2}{\pi^4}
	\sin^4\frac{n\pi}{R+3-L/2}~\left(
	\frac{1}{4}+\frac{\cos^2\frac{n\pi}{R+3-L/2}}
		{R+3-L/2}\right)~\nonumber\\
	&\approx &  -\frac{1}{4}(\lambda' n^2)^2 + 
	\frac{1}{2}(\lambda' n^2)^2~ \frac{4-L}{R} + O(1/R^2) ~, 
\end{eqnarray}
where the expansion is done by taking $R\to\infty$, keeping $L,n$ fixed.
With the standard identification $R=J$, this expression matches the 
$O({\lambda'}^2)$ terms in (\ref{stringfinalexp}), both at $O(1)$ and at 
$O(1/J)$. This is an impressive confirmation of both the two-loop
gauge theory calculation and the string quantization procedure we
have developed.

One would of course like to take this sort of consideration to even higher
orders.  The authors of \cite{Beisert:2003tq} have used integrability
considerations to conjecture an expression for the planar three-loop 
version of the two-loop anomalous dimension (\ref{twoloopgauge}). 
When we generalize that formula to arbitrary level $L$ and take the 
large $J$ limit, we get something which {\it almost} matches the 
$O({\lambda'}^3)$ terms in the string spectrum (\ref{stringfinalexp}); 
the difference is that the factor $3L-12$ multiplying the $O({\lambda'}^3/J)$ 
term in the string formula is replaced by a factor $3L-8$ in the expansion 
of the three-loop gauge theory formula. In short, the two expressions differ by 
a common $O({\lambda'}^3/J)$ shift of all the levels in the supermultiplet 
(while the spacings within the multiplet are the same). One would be tempted
to absorb this shift into a normal-ordering constant but, as explained
in the discussion following eqn.~(\ref{22}), it appears that any further 
normal-ordering freedom is eliminated by supersymmetry. The near-agreement
between the two $O(\lambda'^3)$ expressions is tantalizing, though, at the moment,
we do not see how to make them agree perfectly.

It would be instructive to write down a formula that interpolates between
the string theory energy eqn.~(\ref{stringfinal}), valid for large $J$ and 
finite $\lambda' = \Rhat^4/J^2$, and the gauge theory anomalous dimension 
eqns.~(\ref{deltalevel},\ref{twoloopgaugeL}), valid for small 
$\lambda = g_{YM}^2N_c$ and finite $R$-charge. An additional constraint on 
the interpolation is that anomalous dimensions are predicted to scale 
like $\Delta_n^K \sim 2(n^2\lambda)^{1/4}$ at large values of $\lambda$ 
\cite{Witten:1998qj}.  A simple formula that connects this $\lambda \gg 1$ 
limit with the large-$J$ limit of the string theory is
\begin{eqnarray}
\label{sqrtconjecture}
\left(\Delta_n^K\right)^2 - K^2 =   4\left(1 + \sqrt{K^2+  n^2\lambda}\right)~,
\end{eqnarray}
where $K = J+2-\frac{L}{2}$ is assumed to be large compared to $n$, 
and $\lambda$ is unrestricted. 
For large $J$ (at $L=4$) and fixed $\lambda'$ this gives
\begin{eqnarray}
\Delta_n^J - J & \approx &
	2\sqrt{1+n^2\lambda'} - \frac{2 n^2\lambda'}{J}
	+ \frac{4 n^2\lambda'}{J^2}\sqrt{1+n^2\lambda'}
	+ O(1/J^3)~.
\end{eqnarray}
If the conjecture in eqn.~(\ref{sqrtconjecture}) is correct, a string theory
calculation to $O(1/J^2)$ should yield the coefficient 
$4n^2\lambda'\sqrt{1+n^2\lambda'} $;  
it may be possible to check this by extending the present calculation
to higher orders in perturbation theory.  
While it is possible to generalize
(\ref{sqrtconjecture}) to make contact with the finite-$R$
one and two-loop results in the gauge theory, the results are neither
simple nor unique.


\section{Discussion and Conclusions}
The objective of this study was to compute the leading
finite-radius curvature correction to the Penrose limit of type
IIB string theory in $AdS_5\times S^5$, and thereby verify that
the AdS/CFT correspondence continues to work beyond leading order
in this expansion. The next-to-leading order perturbation to the
plane-wave geometry induces a complicated non-linear interacting
theory on the worldsheet, and carrying out its light-cone
quantization is a rather elaborate enterprise. The 
satisfying end result is that the degeneracy of the BMN spectrum
of string states is lifted in a way that precisely reproduces the
gauge theory operator dimensions at large but finite $R$-charge
and correctly accounts for the extended ${\cal N}=4$
supermultiplet structure dictated by supersymmetry. The success of
this explicit quantization of string theory in a curved RR
background provides, as a side benefit, rather strong evidence for
the correctness of the supercoset manifold construction of the
Green-Schwarz superstring action.

In this paper, we have restricted attention to the set of
physical string states with two worldsheet impurities, or $\Delta_0
= 2$.  It would be a straightforward exercise to extend the
string theory analysis to include higher-impurity states in this
curved background.  In some respects, the corresponding
calculation in the gauge theory should not be difficult; the
lattice Laplacian techniques outlined in section 3 could be a promising
starting point, and one would be able to probe the correspondence
on a different level. 

In an orthogonal direction, it would be interesting to try to
extend this explicit comparison to higher orders in the finite-radius
expansion.  One would be faced with a much more complicated interacting
theory, along with the need to perform sums over physical string states
in order to do degenerate perturbation theory beyond leading order.
This is certainly a worthwhile problem to attack, especially since the
relevant results on the gauge theory side are, by and large, known.
While the complexity of the methods we were compelled to
use in this paper makes the idea of attempting a brute-force extension
to higher orders unappealing, there are several simplifications
that could make such a calculation possible.
What is ultimately needed is some insight
that would enable an exact solution of the worldsheet sigma model,
along the lines of the WZW solution of the $SU(2)$ nonlinear sigma model.

It has been suggested in~\cite{Bena:2003wd} that the complete
light-cone gauge world-sheet action for the type IIB superstring
in the $AdS_5 \times S^5$ background might be integrable. This is
an appealing suggestion that seems very worthwhile to pursue. If
such a program were to be successfully carried out, it would
represent a major advance. The results presented in this paper
could be used as a check by expanding the exact answer in powers
of $1/J$ (or $1/R$) about the Penrose limit and comparing the
first order correction.

\section*{Acknowledgements}

We would like to thank Niklas Beisert, Charlotte Kristjansen, 
Jan Plefka and Matthias Staudacher for useful comments.
CGC would like to thank Caltech and the Gordon Moore Scholars
Program for generous support of a sabbatical visit during which
this work was initiated.
This work was supported in part by US Department of Energy
grant DE-FG02-91ER40671 (Princeton) and by DE-FG03-92-ER40701 (Caltech).




\begin{thebibliography}{99}

\bibitem{Maldacena:1997re}
J.~M.~Maldacena,
Adv.\ Theor.\ Math.\ Phys.\  {\bf 2}, 231 (1998)
[Int.\ J.\ Theor.\ Phys.\  {\bf 38}, 1113 (1999)]
[arXiv:hep-th/9711200].

\bibitem{Witten:1998qj}
E.~Witten,
Adv.\ Theor.\ Math.\ Phys.\  {\bf 2}, 253 (1998)
[arXiv:hep-th/9802150];
S.~S.~Gubser, I.~R.~Klebanov and A.~M.~Polyakov,
Phys.\ Lett.\ B {\bf 428}, 105 (1998)
[arXiv:hep-th/9802109].

\bibitem{Blau:2001ne}
M.~Blau, J.~Figueroa-O'Farrill, C.~Hull and G.~Papadopoulos,
JHEP {\bf 0201}, 047 (2002)
[arXiv:hep-th/0110242].

\bibitem{Metsaev:2001bj}
R.~R.~Metsaev,
Nucl.\ Phys.\ B {\bf 625}, 70 (2002)
[arXiv:hep-th/0112044].

\bibitem{Berenstein:2002jq}
D.~Berenstein, J.~M.~Maldacena and H.~Nastase,
JHEP {\bf 0204}, 013 (2002)
[arXiv:hep-th/0202021].

\bibitem{GenusCounting}
C.~Kristjansen, J.~Plefka, G.~W.~Semenoff and M.~Staudacher,
Nucl.\ Phys.\ B {\bf 643}, 3 (2002)
[arXiv:hep-th/0205033];
N.~R.~Constable, D.~Z.~Freedman, M.~Headrick, S.~Minwalla, L.~Motl, 
A.~Postnikov and W.~Skiba,
JHEP {\bf 0207}, 017 (2002)
[arXiv:hep-th/0205089].

\bibitem{OpDimCalcs}
N.~Beisert, C.~Kristjansen, J.~Plefka, G.~W.~Semenoff and M.~Staudacher,
Nucl.\ Phys.\ B {\bf 650}, 125 (2003)
[arXiv:hep-th/0208178];
N.~R.~Constable, D.~Z.~Freedman, M.~Headrick and S.~Minwalla,
JHEP {\bf 0210}, 068 (2002)
[arXiv:hep-th/0209002];


A.~Santambrogio and D.~Zanon,
Phys.\ Lett.\ B {\bf 545}, 425 (2002)
[arXiv:hep-th/0206079].

\bibitem{BeisertSUSY}
N.~Beisert,
[arXiv:hep-th/0211032].

\bibitem{Beisert:2003tq}
N.~Beisert, C.~Kristjansen and M.~Staudacher,
arXiv:hep-th/0303060;

\bibitem{Parnachev:2002kk}
A.~Parnachev and A.~V.~Ryzhov,
JHEP {\bf 0210}, 066 (2002)
[arXiv:hep-th/0208010].

\bibitem{CGCIanTristan}
C.~G.~Callan, T.~McLoughlin and I.~Swanson,
{\it In preparation. }

\bibitem{Green:1983wt}
M.~B.~Green and J.~H.~Schwarz,
Phys.\ Lett.\ B {\bf 136}, 367 (1984).

\bibitem{MetTseyt}
R.~R.~Metsaev and A.~A.~Tseytlin,
Nucl.\ Phys.\ B {\bf 533}, 109 (1998)
[arXiv:hep-th/9805028].

\bibitem{Kallosh:1998zx}
R.~Kallosh, J.~Rahmfeld and A.~Rajaraman,
JHEP {\bf 9809}, 002 (1998)
[arXiv:hep-th/9805217].

\bibitem{Boerner}
H.~Boerner,
``Representations of Groups''
[North-Holland, 1963] QA171 .B6453 (1970).


\bibitem{QMforDim}
N.~Beisert, C.~Kristjansen, J.~Plefka and M.~Staudacher,
Phys.Lett.B558:229-237,2003, [arXiv:hep-th/0212269] 

\bibitem{minahan}
J.~Minahan and K.~Zarembo, 
JHEP 0303:013,2003, [arXiv:hep-th/0212208].

\bibitem{Bena:2003wd}
I.~Bena, J.~Polchinski and R.~Roiban,
arXiv:hep-th/0305116.

\end{thebibliography}
\end{document}